 \documentclass{aa}				
\usepackage{graphicx}
\usepackage{txfonts}
\usepackage{xtab,booktabs}

\usepackage{natbib,twoopt}
\bibpunct{(}{)}{;}{a}{}{,} 

\usepackage[breaklinks=true]{hyperref} 
\bibpunct{(}{)}{;}{a}{}{,} 
\makeatletter
\newcommandtwoopt{\citeads}[3][][]{\href{http://adsabs.harvard.edu/abs/#3}%
{\def\hyper@linkstart##1##2{}%
\let\hyper@linkend\@empty\citealp[#1][#2]{#3}}}
\newcommandtwoopt{\citepads}[3][][]{\href{http://adsabs.harvard.edu/abs/#3}%
{\def\hyper@linkstart##1##2{}%
\let\hyper@linkend\@empty\citep[#1][#2]{#3}}}
\newcommandtwoopt{\citetads}[3][][]{\href{http://adsabs.harvard.edu/abs/#3}%
{\def\hyper@linkstart##1##2{}%
\let\hyper@linkend\@empty\citet[#1][#2]{#3}}}
\newcommandtwoopt{\citeyearads}[3][][]%
{\href{http://adsabs.harvard.edu/abs/#3}
{\def\hyper@linkstart##1##2{}%
\let\hyper@linkend\@empty\citeyear[#1][#2]{#3}}}
\newcommand\vv{{\mathrm v}  }
\makeatother

\begin{document}

\title{Alpha Virginis: line-profile variations and orbital elements}
\titlerunning{Orbital elements of Alpha Vir.}

\author{David Harrington  \inst{1,} \inst{2,} \inst{3} \and Gloria Koenigsberger \inst{4} \and Enrique Olgu\'{\i}n \inst{7} \and Ilya Ilyin \inst{5} \and Svetlana V. Berdyugina \inst{1,} \inst{2} \and Bruno Lara \inst{7} \and Edmundo Moreno \inst{6}}

\institute{Kiepenheuer-Institut f\"{u}r Sonnenphysik, Sch\"{o}neckstr. 6, D-79104 Freiburg, Germany \\
\and Institute for Astronomy, University of Hawaii, 2680 Woodlawn Drive, Honolulu, HI, 96822, USA  \\
\and Applied Research Labs, University of Hawaii, 2800 Woodlawn Drive, Honolulu, HI, 96822, USA \\
\and Instituto de Ciencias F\'{\i}sicas, Universidad Nacional Aut\'onoma de M\'exico, Ave. Universidad S/N, Cuernavaca, Morelos, 62210, M\'exico \\
\and Leibniz-Institut f\"{u}r Astrophysik Potsdam (AIP), An der Sternwarte 16, 14482 Potsdam, Germany \\
\and Instituto de Astronom\'{\i}a, Universidad Nacional Aut\'onoma de M\'exico, Apdo. Postal 70-264, M\'exico, D.F. 04510, M\'exico \\
\and Centro de Investigaci\'on en Ciencias, Universidad Aut\'onoma del Estado de Morelos, Cuernavaca, 62210, Mexico}

\date{Submitted April, 2015}


\abstract{ {\it Context:} Alpha Virginis (Spica) is a B-type binary system whose proximity and brightness allow detailed investigations of the internal structure and evolution of stars undergoing time-variable tidal interactions.  Previous studies have led to the conclusion that the internal structure of Spica's primary star may be more centrally condensed than predicted by theoretical models of single stars, raising the possibility that the interactions could lead to effects that are currently neglected in structure and evolution calculations. The key parameters in confirming this result are the values of the orbital eccentricity $e$, the apsidal period $U$, and the primary star's radius, $R_1$.  \\
{\it Aims:} The aim of this paper is  to analyze the impact that  Spica's line profile variability has 
on the derivation of its orbital elements and to explore the use of the variability for constraining $R_1$. \\
{\it Methods:} We use high signal-to-noise and high spectral resolution observations obtained in 2000, 2008, and 2013 to derive the orbital elements from fits to the radial velocity curves. We produce synthetic 
line profiles using an {\it ab initio} tidal interaction model. \\
{\it Results:} The general variations in the line profiles can be understood in terms of the tidal flows, whose large-scale structure is relatively fixed in the rotating binary system reference frame. Fits to the radial velocity curves yield $e$=0.108$\pm$0.014.  However, the  analogous RV curves from theoretical line profiles indicate that the distortion in the lines causes the fitted value of $e$  to depend on the argument of periastron; i.e., on the epoch of observation. As a result, the actual value of $e$ may be as high as 0.125. We find that $U$=117.9$\pm$1.8, which is in agreement with previous determinations. Using the value $R_1=6.8 R_\odot$ derived by Palate et al. (2013) the value of the observational internal structure constant $k_{2,obs}$ is consistent with theory.  We confirm the presence of variability in the line profiles of the secondary star.}

\keywords{stars: individual ($\alpha$ Vir) ---
          stars: binaries: fundamental parameters  ---
          stars: binaries: tides ---
          stars: spectral variations }

\maketitle

\section{Introduction}

	A fundamental  question in stellar astrophysics is whether the internal structure and the evolution of binary stars is the same  as in single stars of equivalent mass, chemical composition, and rotation. Binary stars provide the only means of directly determining stellar mass, and their masses are generally adopted as representative of single stars. Reasons exist, however, to question whether the generally assumed equivalence between single and binary stars  is a valid assumption.  For example,  tidal shear energy dissipation in asynchronous binaries may lead to systematic differences in the internal temperature structure \citep{1968Ap&SS...1..411K, 1975ApJ...202L.135P}.  The differential velocity structure produced by the tidal interactions could lead to internal mixing rates that are significantly different from those derived from the simple velocity structure generally assumed for single stars \citep{2013EAS....64..339K}.  It is in the context of these issues that the importance of the $\alpha$ Virginis ({\it Spica}) binary system cannot be overstated.

	The system $\alpha$ Virginis (Spica, HD 116658) consists of two B-type stars in a short-period ($\sim$4 d) eccentric  orbit, and it is one of the closest, brightest and most extensively studied of the known double-lined spectroscopic binaries. It was resolved interferometrically by \citet{HerbisonEvans:1971th}, hereafter HE71, and their results, combined with results obtained from radial velocity curves, yielded the orbital elements and stellar parameters that are  listed in Table \ref{spicaprop}. \citet{2007IAUS..240..271A} focused on the value of the period of rotation of the line of apsides ($U$), a parameter that is related to the internal mass distribution, and confirmed the earlier findings that the observational internal structure constant ($k_{2,obs}$) is too small compared with what would be expected from theory \citep{1973ApJ...180..517M, 1974ApJ...192..417O, 1993A&A...277..487C, 2002A&A...388..518C, 2003A&A...399.1115C}. The nature of the discrepancy between the observational and theoretical values of $k_2$ implies that the primary star is more centrally condensed than predicted from the models and, if true, would lend evidence supporting the idea that close binary stars are different from equivalent single stars.  Unfortunately, however, the uncertainties in Spica's parameters are still too large to allow a firm conclusion to be reached.  Particularly troubling is the very large uncertainty associated with the radius of the primary star \citep{2015AAS...22534520A}.

        The B1 III-IV primary star in Spica is well known to display strong photospheric line-profile variability which, if thoroughly understood, could yield stellar structure information.  Among the notable features of the profile variability is the presence of traveling ``bumps'' that migrate from the blue to the red wing of the absorption, most clearly seen in the weaker lines of OII and Si III \citep{1982PASP...94..143W, 1983HvaOB...7..231F, Smith:1985dg}.  Based on the explanation put forth for  similar features observed in $\zeta$ Oph \citep{Vogt:1983ic}, Spica's primary star has been classified as a $\beta$ Cep-type star; i.e., it is believed to be be undergoing non-radial pulsations (NRP).

\begin{table}
\begin{center}
\begin{small}
\caption{Spica parameters \label{spicaprop}}
\begin{tabular}{lcc}
\hline
\hline
{\bf Parameter  } 		& {\bf Literature}		&{\bf This paper}		\\ 
\hline
\hline
$P/days$                        & 4.014597 $^1$                 & 4.014597 (fixed)          \\      
$T_0/JD$                        & 2440678.09 $^1$               & 2440678.09 (fixed)         \\
e                               & 0.146 $^1$                    & 0.108$\pm$0.014 (s.d.)$^8$     \\   
                                &                               &0.125$\pm$0.016$^9$      \\
$i/deg$                         & 66$\pm$2 $^1$                 & 66 (fixed)                    \\
$\omega_{per}/deg$ at $T_0$     & 138 $\pm15$    $^1$           & 142 (interpolated)  \\
$U/yrs$                         & 124$\pm$11    $^1$            & 117.9 $\pm$1.8         \\
                                & 118.9$\pm$1.3  $^6$           & ....                 \\
$\vv_0/km~s^{-1}$ for $m_1$     & 0$\pm$2        $^4$           & ....                  \\
$\vv_0/km~s^{-1}$ for $m_2$     &  2$\pm$3        $^4$          & ....                          \\
$\vv_0/km~s^{-1}$               &  ....                         & 0.0 $\pm$3.1 (s.d.)          \\
$K_1/km~s^{-1}$                 & 124$\pm$4      $^4$           & 121.1$\pm$2.8         \\ 
$K_2/km~s^{-1}$                 & l97$\pm$8      $^4$           & 189.8$\pm$2.7      \\
$m_1$/$m_2$                     & 1.60$\pm$0.3 $^1$             & 1.55$\pm$0.09 (s.d.)  \\
                                & 1.59$\pm$0.03      $^4$       &   ....                 \\
$m_1/M_\odot$	        	& 10.9$\pm$0.9  $^1$ 		& 10.0 $\pm$0.3(s.d.)  \\
$m_2/M_\odot$                   &  6.8$\pm$0.7   $^1$     	& 6.4 $\pm$0.2 (s.d.)   \\
R$_1$ (R$_\odot$)       	& 8.1$\pm$0.5 $^1$     		&   ....	\\ 
                                & 7.6 $\pm$0.2 $^7$             &  ....          \\
                       		& 6.84 $^5$			& ....  		\\
R$_2$ (R$_\odot$)       	& 3.64 $^5$			&  ....                       \\
$\vv_1 sin i/km~s^{-1}$ 	& 161$\pm$2{\bf $^2$} 		& ....		\\
$\vv_2 sin i/km~s^{-1}$      	&  70$\pm$5{\bf $^3$ }	        & ....			\\
\hline
\hline
\end{tabular}
\end{small}
\end{center}
{\bf Notes:} {\small  Parameters from:  $^1$\citet{HerbisonEvans:1971th}, $^2$\citet{Smith:1985dg}, $^3$\citet{2000PhDT.......148R}, $^4$ Shobbrook et al. (1972), $^5$ Palate et al. (2013), $^6$ \citet{2015AAS...22534520A}, $^7$ \citet{Sterken:1986tp}, $^8$Average over the 3 data sets analyzed in this paper, $^9$Adopting the hypothesis that $e$ dependes on $\omega_{per}$.}
\end{table}

	This scenario was reinforced by the results of \citet{1972MNRAS.156..165S} who found a 4.18 hr period in both the photometric light curve and the radial velocities. However, the amplitude of these variations declined between the two epochs of observation that they reported, and could no longer  be found a few years later as referenced by a private communication from Shobbroock (1984) in \citet{Smith:1985dg}.  Meanwhile, \citet{1974ApJ...192...81D} had detected a 6.6 hr period which was re-discovered by \citet{Smith:1985dg} who associated it with an $l$=8 NRP mode having $P_{NRP}$= 6.52$\pm$0.08 hr.  

 Although NRPs provide a tool for deriving information on the internal stellar structure, both stars in the Spica system rotate asynchronously.  Because of the close orbital separation, significant tidal perturbations are present which preclude the interpretation of the variability in terms of NRPs alone.  Indeed,  \citet{Smith:1985dg} found two  modes additional to the $l$=8 mode which he concluded were caused by the tidal interaction of the primary star (henceforth $m_1$) with the companion (henceforth $m_2$).  The first of these is due to the equilibrium tidal distortion and \citet{Smith:1985dg} described it as the ``spectroscopic equivalent of the photometric ellipsoidal variability'' equilibrium tidal distortion. This mode is associated with a change in the general shape of the line profile and has a period $P_{orb}/2$, where $P_{orb}$ is the orbital period.  \citet{Smith:1985dg} described the second tidally-induced mode as a ``quasi-toroidal'' mode, deduced from the presence of a strong absorption ``spike'' that remains in a quasi-stationary position for $\sim$2 hour intervals and is located ``just inside (alternately) the red or blue edges of the profile''. The repeatability timescale for the appearance of this spike was found to be $\sim$8 hours; i.e., P/12.  

	\citet{Harrington:2009ka} theoretically reproduced the general behavior of the traveling ``bumps'' and the blue/red ``spike''  using a one-layer model which provides a solution from first principles of the equations of motion governing the tidal interaction \citep{1999RMxAA..35..157M, Moreno:2005cq}.  The basic conclusion of this investigation is that the observed line-profile variability is caused largely by the horizontal velocity field on the stellar surface. These horizontal motions, described in \citet{Harrington:2009ka} as {\it tidal flows}, correspond in principle to the tidally-induced modes described by  \citet{Smith:1985bd}.  Because the tidal effects are extremely sensitive to the ratio of stellar radius to orbital separation, this opens the interesting possibility of using the line profile variability as an independent constraint on the stellar radius.  Before this can be achieved, however, the impact of the line-profile variability on the measured radial velocities (RVs) needs to be assessed.  This can only be achieved through a detailed analysis of the variability in very high quality observational data.

	In this paper, we revisit Spica's orbital parameters by combining three sets of spectroscopic observations obtained in 2000, 2008 and 2013 to address the manner in which the line profile variability may influence the results.  Section \ref{sec:observations} describes our observational data and method of RV measurements, and Section 4 contains the description of the observed line profile variability.  In section 5 we illustrate with theoretical calculations the manner in which the tidal flows lead to the appearance of the spikes in the line profiles and the manner in which the changing orientation of the elliptical orbit over time leads to a variable determination of $e$.  In Section 6 we summarize the conclusions.

\begin{figure} 
\includegraphics[width=0.98\linewidth, angle=0]{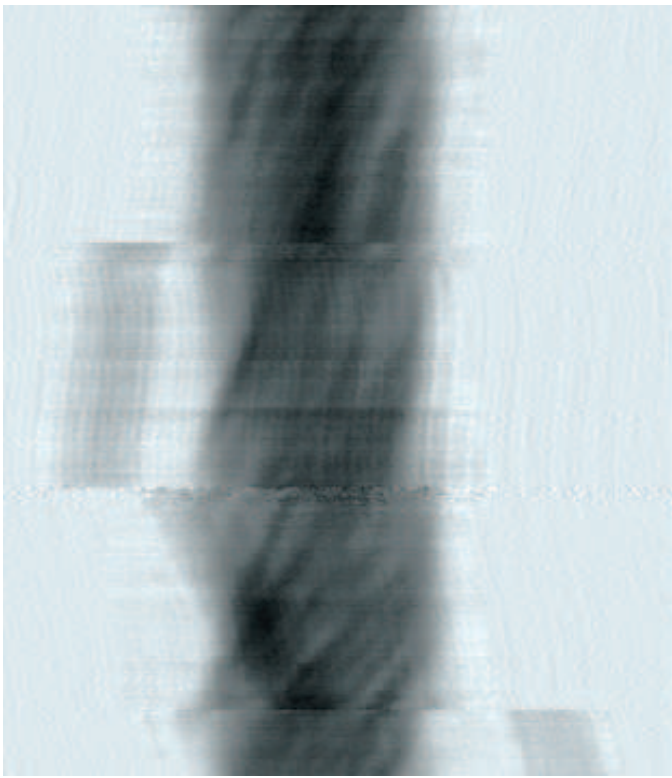}
\caption{\label{greyscale} Grayscale montage of the 4552 \AA\ Si III line recorded with the SOFIN spectrograph. The abscissa corresponds to the radial velocity in the frame of reference centered in $m_1$ and the observations are stacked from bottom to top according to sequence number of the observation starting with the first spectrum obtained on 19 April. }
\end{figure}

\begin{figure} 
\includegraphics[width=0.98\linewidth]{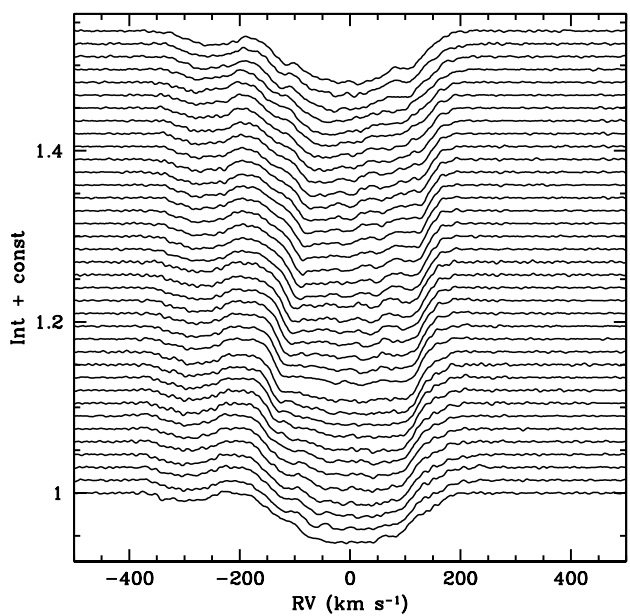}
\caption{ Line profiles of Si III 4551 obtained on JD 2451655  corrected for the  orbital motion of $m_1$ using the measured RVs from Table \ref{table_SOFIN_RVs} (see Appendix). These spectra cover a timespan of 8.67 {\rm hr} and correspond to the orbital phase range 0.34 -- 0.43. The $m_2$ absorption is centered on $-$280 km s$^{-1}$ in this reference frame. Each profile is shifted along the ordinate by a fixed amount for display purposes, with time running from bottom to top.
\label{averages_block2}
}
\end{figure}

\begin{table}
\begin{center}
\begin{small}
\caption{Summary of observations \label{table_observations}}
\begin{tabular}{lccc}
\hline
\hline
{\bf   }          & {\bf 2000}  &{\bf 2008} & {\bf 2013}    \\
\hline
\hline
Dates             &April 19-22  & March 15-28 &May 18-30    \\
                  &             &             &June 19-29     \\
Num. Spectra      &  577        &  13         & 17            \\
Wavelength (\AA)  &4536-4566    &3600-10000   & 3600 --10000  \\
S/N               &$\sim$500    &1000-2000  &1000-2000      \\
Orbital Phases$^1$&0.09-0.18    &0.09-0.15    & 0.06-0.14     \\
                  &0.34-0.43    &0.36-0.40    & 0.31-0.35     \\
                  &0.59-0.66    &0.63-0.67    & ....          \\
                  &0.89-0.94    &0.85-0.88    & 0.82-0.87     \\
\hline
\hline
\end{tabular}
\end{small}
\end{center}
{\bf Notes:} {\small $^1$Orbital phases are measured from periastron}
\end{table}

\section{Observations}
\label{sec:observations}
Three sets of observations were acquired between 2000 and 2013, the first using the Nordic Telescope ({\it NOT}) and the second two with the Canada France Hawaii Telescope ({\it CFHT}). The general characteristics of these data sets are summarized in Table \ref{table_observations} and are discussed in greater detail below.

In addition, we used the RV measurements obtained in 2000 that are listed in \citet{2000PhDT.......148R} which complement the phase coverage of the NOT data and, in addition, which illustrate the challenges for deconvolution methods intended to separate the primary and secondary line profiles.
 
\subsection{SOFIN}

        The Nordic Optical Telescope (NOT) was used to acquire a set of 621 spectra of Spica on 4 consecutive nights in 2000, April 19--22. The SOFIN echelle spectrograph was used in a high resolution mode (R=80,000) sampled at 0.029 \AA\  per pixel for a 1-pixel spectral sampling of 163,000 or 1.8 km/s per pixel. The exposure times were set to 110$s$ to 130$s$ following the airmass of the target. The conditions were not entirely photometric and after discarding the very low S/N exposures we are left with 577 observations used in the RV analysis.

The RV stability of this instrument is limited by the slit error: a slow motion of the star on the slit due to guiding and atmospheric dispersion. Typically the RV error is in a range of 50-100 m/s. Any long term temporal variations in the Cassegrain mounted spectrograph due to change of the environment conditions and flexure of the spectrograph were compensated by taking the ThAr lamp exposures every 20 minutes. The two ThAr exposures before and after were combined and used for the wavelength calibration of the bracketed target exposures. The analyzed spectral order is centered on the Si III 4552 \AA\ line. 

The response of the echelle orders is complicated and varies over short time intervals, thus special care is required for the order rectification.  A semi-automatic routine was written to trace the continuum level by performing a boxcar average over 5 points along the continuum and interpolating a straight line across the absorption lines.  The largest uncertainties in this approach are in: a) choosing the location where the absorption line wings reach the continuum; and b) the linear interpolation across the line. The first of these has a minor effect on the $m_1$ measurements because of its generally very pronounced line wings.  Its effect on the $m_2$ lines is potentially greater due to their relative weakness, so particular care was needed for the location of its wings. The second of these does not represent a significant issue since visual inspection of the un-normalized spectra confirms that in the great majority of cases the continuum around the absorption lines has a linear shape. The normalization was performed on each of the spectra individually, which provided the first pass of the original 621 exposures, in which poor quality spectra  were eliminated.

The signal-to-noise ratio (SNR) was measured on the individual normalized spectra yielding in general S/N$\sim$500 in the vicinity of the Si III absorption lines, and falling to 300-400 near the edges of the echelle order. Spectra with the poorest S/N were excluded from the RV analysis, leaving 577 usable spectra, and a further dozen spectra were excluded for the line-profile variability analysis.

 The orbital phase coverage and other characteristics of the observations are listed in Table \ref{table_observations}, and Fig. \ref{greyscale} displays the 570 spectra, stacked vertically in order of increasing phase after applying the correction for the orbital motion of $m_1$ that is obtained in Section \ref{sec:RVs}. 

\subsection{ESPaDOns}

        Two sets of ESPaDOns observations were obtained in 2008 and 2013 in queue mode at the Canada France Hawaii  (CFHT) 3.6m telescope with the ESPaDOns spectropolarimeter at a nominal spectral resolution of R=68,000. ESPaDOnS is a fiber-fed cross-dispersed echelle covering 370nm to 1048nm in a single exposure over 40 spectral orders on the EEV1 detector.  The first set, obtained in 2008, March 15--28, is described in \citet{Harrington:2009ka}. The second set, obtained in 2013, May 18--30 and June 21--27,  consists of 17 spectra in the orbital phase intervals 0.1--0.35 and 0.81--0.88, where the fastests radial velocities occur and the absorption lines from each of the two stars are well separated. Phase $\phi$=0 corresponds to periastron passage.  A subset of 6 spectra obtained within a 0.8 hr timespan is included in the second set.  

        The instrument has a dedicated reduction package, Libre-Esprit that automatically processes the data. This script does typical flat fielding, bias subtraction, wavelength calibration using both calibration lamp and Fabry-Perot frames and optimal spectral extraction \citep{Donati:1997wj}. We have written additional post-processing scripts to combine orders and bin to lower effective spectral sampling, increasing the effective signal to noise ratios (SNRs) by factors of typically 2 to 3 at shorter wavelengths with reduced sampling \citep{Harrington:2009ka}.  The two data sets consist of spectra with 4.4 km/s resolution at 4600 \AA. (R=68,000). With our post-processing scripts, we achieve effective signal-to-noise\footnote{Excluding systematics such as echelle inter-order overlap and CCD artifacts} of $\sim$1000--3000 at 4600 \AA\ with a sampling of 0.12 \AA\  per pixel which is equivalent to 7.8 km/s per pixel. Extreme care is required in treating the wavelength regions where the echelle orders overlap to avoid degrading the  S/N and/or the appearance of spurious features.  The phase coverage and other characteristics of this data set are listed in Table \ref{table_observations}.

\begin{figure} 
\begin{center}
\includegraphics[width=0.95\linewidth]{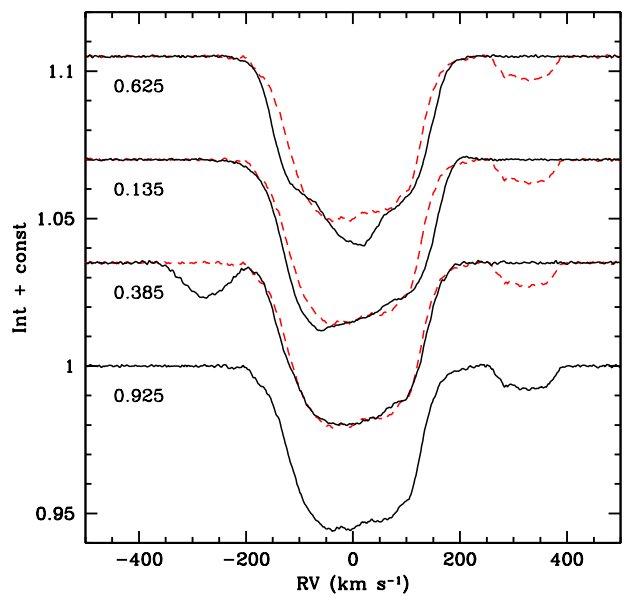}
\caption{\label{compare_averages}  Average profiles for each night in the 2000 data plotted on a velocity scale corrected for the measured RVs of $m_1$. A constant vertical shift is applied for clarity in the figure.  The average orbital phase for the night is shown. The spectrum at $<\phi>$=0.925 is re-plotted  (dashes) for comparison with the others, and shows that the line-profile near conjunctions ($<\phi>$=0.135 and 0.625) is broader than near elongations ($<\phi>$=0.925, 0.385).
}
\end{center}
\end{figure}

\section{Line profile variability}

The variations in the line profiles observed in $m_1$ are characterized by:  a)  appearance/disappearance of superposed quasi-stationary absorption dips and spikes, b) the presence of traveling ``bumps'' that migrate from the blue to the red wing, and c) variations of the extent and slope of absorption line wings.  These characteristics combine to cause the overall line shape to change from one similar to a paraboloid to one having a very flat core with steep wings.  These changes are evident in the  weak absorption lines such as those produced in  O II and Si III transitions \citep{1982PASP...94..143W, 1983HvaOB...7..231F, Smith:1985dg,  Harrington:2009ka}, and are illustrated in the sequence of SOFIN profiles of Si III 4552  shown in Fig. \ref{averages_block2}. This sequence was obtained during the orbital phase interval 0.34-0.43,  when the $m_1$ and $m_2$ lines are well resolved, and covers a timespan of 0.8.67 {\rm hr}.  Analogous plots for the other 3 nights of observation are presented in the Appendix in Figs. \ref{averages_block13} and \ref{averages_block4}.

\subsection{Broader wings around conjunction}

\begin{figure} 
\begin{center}
\hbox{
\hspace{-1.0em}
\includegraphics[width=1.03\linewidth, angle=0]{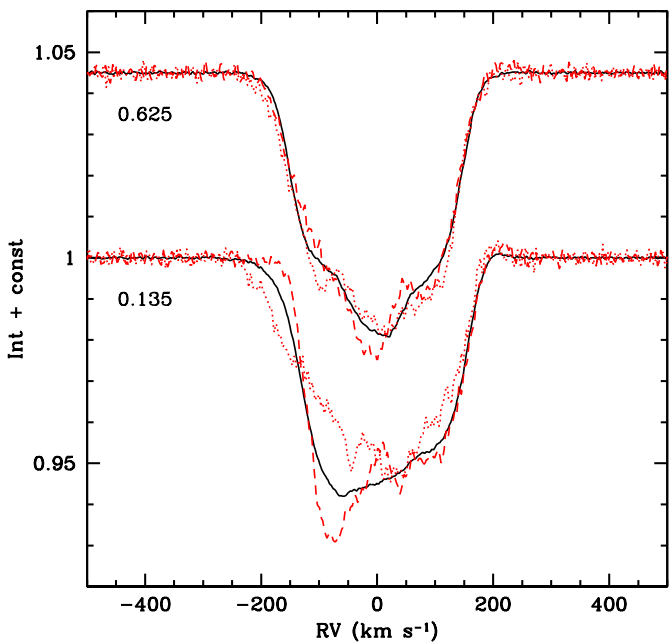}
}
\caption{\label{compare_averages2} Average line profile at both conjunctions (continuous line) is compared with the first (dash) and the last (dots) of the individual 5-spectrum averages of the same night, illustrating that the spectra at these orbital phases are all broader than near elongations.  The significantly different profile at $<\phi>$=0.135 is caused by the presence of $m_2$ having moved from near the core of $m_1$'s absorption to its wing over the $\sim$7 hours that these observations lasted. The rapid motion is due to the fact that this phase is just after
periastron.
}
\end{center}
\end{figure}

A new feature of the Spica line profiles  emerges from a comparison of the average $m_1$ profiles for each night: Fig. \ref{compare_averages} shows that the half-width of the line around conjunctions is on average 15 $km~s^{-1}$ broader than at elongations.  This excess width is  not due to the blending with $m_2$'s line, which lies mostly in the core of $m_1$'s line.  It is a constant feature over the orbital phase interval over which the averages are constructed, as shown in Fig. \ref{compare_averages2} where these averages over the night are compared with the first and last spectra of the night. Harrington et al. (2009) showed that the tidal flows can attain high speeds and that they strongly affect the shape of the line-wings. This  suggests that the different widths may be associated with different flow speeds when the system is viewed at conjunctions as opposed to elongations.

\subsection{Parabolic and ``boxy'' shapes}

\begin{figure} 
\includegraphics[width=1.05\linewidth]{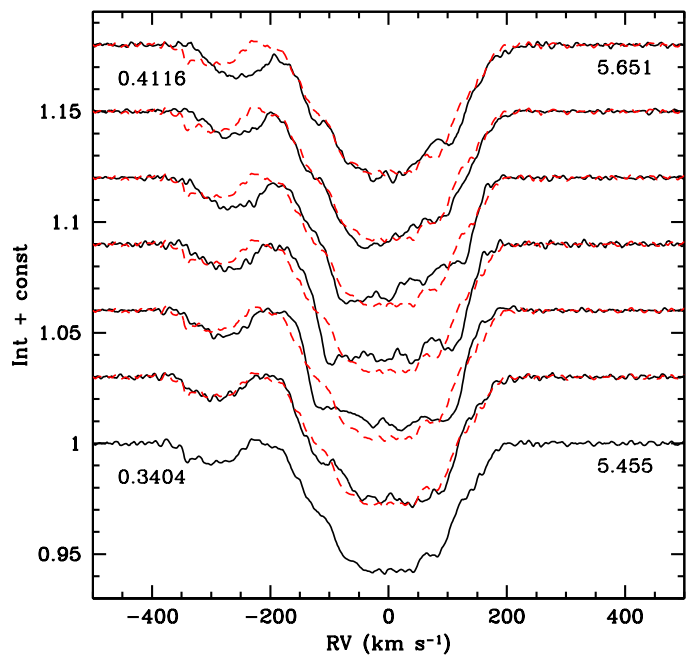}
\caption{Over a timespan of 4.7 hrs the shape of $m_1$'s line cycles from paraboloid, to flat-bottomed, to sloping upward towards the blue and then approaching once again the paraboloid shape.  Time runs from bottom to top in units of days and is indicated on the right side. The corresponding orbital phase is listed on the left. The profile shown with dashes is always the same one ($\phi$=0.34). The profile of $m_2$ appears here at $-$280 $km~s^{-1}$. 
\label{trends_block2}}
\end{figure}

One of the trends in the overall line profile shape is to transition from a paraboloid-looking profile with a rounded core, to a ``boxy'' shape in which the core is very flat and at least one of the wings  is very steep.  This is seen, for example, on JD 2451655 (Fig. \ref{averages_block2}) where a feature that appears near the continuum level on the blue wing propagates redward for $\sim$4.7 {\rm hrs}, after which time a new such feature appears.  As this occurs,  the shape of $m_1$'s line goes from paraboloid, to one with a flat-core that then gradually slopes upward towards the blue, to then once again become a paraboloid. A more detailed look at this phenomenon is presented in Fig. \ref{trends_block2}, where the first spectrum is used as a template for comparison with the subsequent ones.  The transition seems to be associated with the appearance, growth and migration of  blue and red spikes, accompanied by a number of smaller ``bumps''. 

\subsection{Spikes and ``bumps''}

\begin{figure} 
\includegraphics[width=1.05\linewidth]{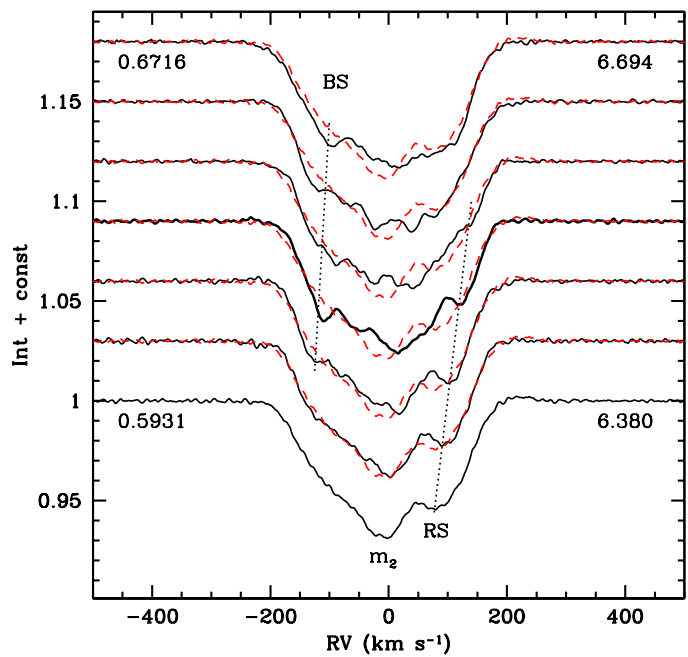}
\caption{Sequence of profiles over a 7.5 hr timespan that includes the time of conjunction when $m_2$ is closer to the observer (bold-line). $m_2$ can be gleaned to march redward through the line core of $m_1$ over the first 5 spectra (from bottom to top) and then its presence is less obvious.  Noteworthy is the presence of a red spike (RS) that propagates redward and then seems to vanish shortly after the blue spike (BS) appears around the time of conjunction. The BS  remains in view until the end of the sequence. \citet{Smith:1985dg} also noted that the spikes ``seem to be strongest when the hemisphere of $m_1$ facing $m_2$ is also facing the Earth''.
\label{trends_block4}}
\end{figure}

\citet{Smith:1985dg}  noted the repetitive nature of the appearance and disappearance of the spikes, and particularly significant is his comment that the blue and red spikes can appear simultaneously and ``seem to be strongest when the hemisphere of $m_1$ facing $m_2$ is also facing the Earth''.  We find the same to be true in our observations. This is illustrated in the sequence of profiles obtained over a 7.5 hr timespan on JD  2451656, shown in Fig. \ref{trends_block4}, which includes the time mentioned by \citet{Smith:1985dg} in the statement above,  coinciding with the conjunction when $m_2$ is closer to the observer.  At the start of the sequence, the red spike is very strong  and located at $\sim$75 {\rm km~s$^{-1}$}.  It gradually migrates towards the red and decreases in strength. Just before conjunction, the blue spike appears at $-$120 {\rm km~s$^{-1}$} and both spikes coexist for a short period of time, before the red spike is not longer visible.  The fact that we observe the same phenomenon as did \citet{Smith:1985dg} at the same geometrical configuration of the binary system supports the idea that these variations are orbital-phase locked and are caused by the tidal flow structure on $m_1$'s surface.  Note, however, that ``phase-locked'' in this context may refer to time measured with respect to conjuntions rather than with respect to periastron passage.

The above being said, it is important to note that although some of the properties of the line profile variability can be seen to repeat from one orbital cycle to another, the smaller details are not strictly periodic on orbital timescales.  For example, the set of 6 spectra that were obtained with ESPaDOnS within a $\sim$47 minute timespan on 2013 June 23 show a prominent ``bump'' that migrates from the blue to the red at a rate of $\sim$20 {\rm km~s$^{-1}$~per~second}, and a similar feature is present  precisely one orbital cycle earlier, on June 19.  However, the ``red spike'' which is strong on June 19 is only marginally evident on June 23 (see Fig. \ref{2013Jun23_blue_spike} in the Appendix).  Hence, the behavior of the smaller features (``bumps'') in the line profiles is most likely governed by internal modes of pulsation and local hydrodynamical effects that lead to time scales that are independent of the orbital timescale.

\begin{figure} 
\includegraphics[width=1.03\linewidth]{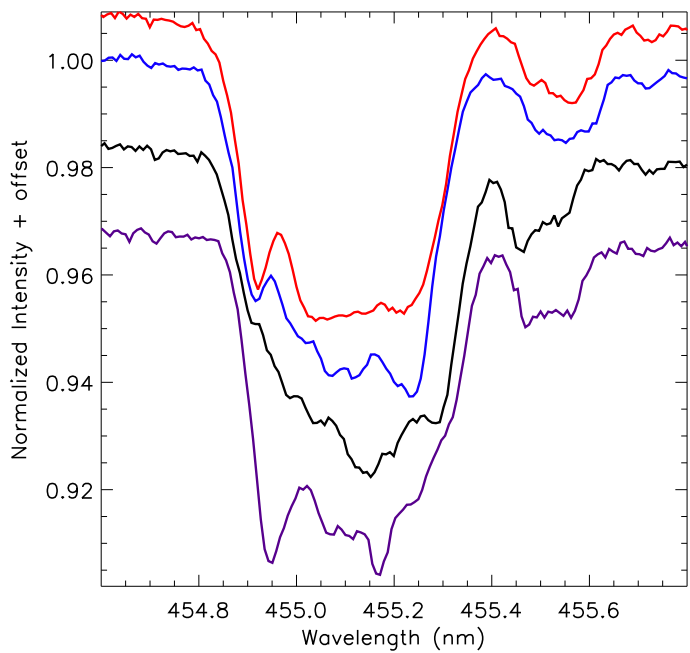}
\caption{Si III 4552 \AA\ absorption arising in $m_1$ (left) and in $m_2$ (right) in spectra obtained in 2013 in the phase interval 0.82--0.88.  Note the varying shape of both the $m_1$ and $m_2$ absorption lines. From bottom to top: May 30 ($\phi$=0.854), May 18 ($\phi$=0.877), June 19 ($\phi$=0.822) and June 23 ($\phi$=0.826).
\label{variability_m2}}
\end{figure}

\subsection{The line profiles of $m_2$ are also variable}

We end this section with a few comments on the line profiles of $m_2$. \citet{2000PhDT.......148R} noted that  its lines are systematically weaker at one elongation with respecto to the other, but that this is an epoch-dependent phenomenon.  Our data confirm that $m_2$'s line profiles are indeed variable, having a paraboloid shape when approaching the observer and a flat-bottom shape at the opposite elongation (Fig. \ref{compare_averages}), which makes it appear weaker. Furthermore, it also seems to have blue and red spikes at similar times as these spikes appear in $m_1$'s profiles (Fig. \ref{variability_m2}). From a qualitative standpoint, this is not unexpected since  $m_2$ is also in non-synchronous rotation and therefore is also expected to present tidal flow phenomena and possibly also NRPs.

\section{Radial velocities and orbital elements}
\label{sec:RVs}

We have previously pointed out \citep{Harrington:2009ka} that the determination of Spica's radial velocity curve is not a straightforward process due to the deformed shape of the absorption line profiles and their variability.  Specifically, there is no unambiguous way to determine the line centroid which, in non-perturbed atmospheres,  provides the orbital motion.  The line shape cannot even remotely be approximated with  Gaussian or Voigt functions, a procedure usually employed when measuring RVs. Thus, the only consistant method is to use the intensity-averaged centroid as the measure of the line RVs. This method, however,  yields values that can differ by as much as 16 km s$^{-1}$ depending on whether the line is measured at continuum level or closer to the core \citep{Harrington:2009ka}, so extreme care is required to perform the measurements consistently at continuum level.

Measurements near elongations are straightforward, since both stars are generally well resolved. The problem arises for orbital phases in which the lines overlap.  Under classical conditions (i.e., the line profiles remain stable over the orbital cycle), a deconvolution technique is applicable.  However, the application at this time of such techniques to Spica is doomed to failure due to the profile variability in both stars.  In particular, the line profiles of $m_1$ are prone to present dips near the line core that mimic the line profile of
$m_2$.  When the lines overlap, this dip combines with the $m_2$ absorption making it nearly impossible to obtain the RV of $m_2$ in an unambiguous manner. We illustrate the problem in Fig. \ref{same_profiles}, where at orbital phase $\phi$=0.925 the dip near the $m_1$ core is seen at the same time as  $m_2$'s line is at $+$300 km s$^{-1}$.  At the other phases plotted in this figure (near conjunctions), the dip overlaps the $m_2$ absorption enhancing its apparent strength.  The actual shape and location of the dip is unknown and, as noted above, the line profiles of $m_2$ are also variable.  Hence, attempting to deconvolve the RVs at these orbital phases with classical techniques is likely to yield spurious RV values.

In this section, our approach is the following: 1) We  measure the centroid of $m_1$ on all the exposures using the intensity-weighted centroid method, and that of $m_2$ only when it can be clearly isolated from the absorption of $m_1$.  2) We adopt the measured RVs as the best approximation to the projected orbital motion of $m_1$ and plot the line profiles on a velocity scale that is Doppler-shifted correspondingly.  3) We assess the impact that the line-profile variability and blending may have on the measured RVs and determine whether  the values obtained for $m_2$ around conjunctions are trustworthy. 4) The orbital elements are dervied from the fit to the final list of RVs.

\begin{figure} 
\begin{center}
\hbox{
\hspace{-0.0em}
\includegraphics[width=1.03\linewidth, angle=0]{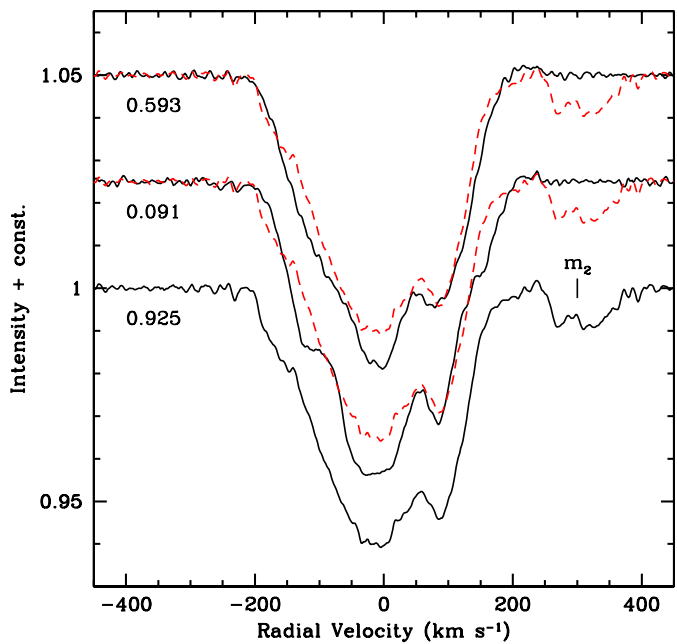}
}
\caption{\label{same_profiles} SOFIN line profiles just before (0.925) and just after (0.091) periastron, and just after  apastron (0.593) have nearly the same shape.  Note the deep central dip just before periastron which, when observed at phases when $m_1$ and $m_2$ are not resolved, mimics the $m_2$ absorption.  Note also that these phases are close to conjunctions.  The profile at 0.925 is superposed on the others with dashes. 
\label{same_profiles}
}
\end{center}
\end{figure}

\subsection{RV measurements}

The RVs  were obtained by measuring the intensity-weighted line centroids using IRAF.\footnote{The Image Reduction and Analysis Facility (IRAF) is distributed by the National Optical Astronomy Observatory, which is operated by the Association of Universities for Research in Astronomy (AURA) under a cooperative agreement with the National Science Foundation.}  In the SOFIN  and ESPaDOnS data sets, we measured Si III 4552.62 \AA. In the ESPaDOnS sets we also measured He I 6678.15 \AA,  O II 4906.83 \AA and O II 4661.63 \AA. These lines were all chosen  because they are sufficiently isolated from neighboring atomic transitions to avoid blending effects.

The measurement uncertainties for the Si III and He I lines are $\sim$1 km s$^{-1}$ for $m_1$ and $\sim$2 km s$^{-1}$ for $m_2$ when both components are clearly resolved. For the weaker O II lines the uncertainty is $\sim$3 km s$^{-1}$, driven primarily by the choice of the continuum level.  These uncertainties do not take into account systematic effects which are produced by the line-profile variability.

In what follows, we focus on the RV measurements of the SOFIN set since the 2008 data have been discussed previously \citep{Harrington:2009ka} and the 2013 data were obtained only near elongations.

During two of the 4 nights of observations, the $m_1$ and $m_2$ components are well separated and do not require further discussion.  This is not the case of the other two nights, when these components overlap completely. During the first, $m_2$ is near the core of $m_1$ at the beginning of the night and in the blue wing by the end of the night. Its location can actually be traced on the grayscale representation in Fig. \ref{greyscale} and its rapid motion is due to the fact that this night corresponds to orbital phases just after periastron passage. We performed numerous tests to  measure the RVs of $m_2$ around conjunction, aided by the fact that the long sequence of observations allows one to follow the motion of $m_2$ through the core of $m_1$'s absorption and into the blue wing. However, these measurements are  significantly more uncertain than those near elongations.

During the other night, $m_2$ starts out again near the core of $m_1$ and gives the impression that it is moving rapidly towards the red wing over approximately one third of the night, after which time it is no longer visible.  We believe that its good visibility on the grayscale plot is due to a coincide of its location with that of a migrating ``bump''. Once the ``bump'' migrates out of $m_2$'s wavelength range, $m_2$'s presence is no longer as evident.  We thus consider the RV measurements of $m_2$ in this orbital phase range to be highly unreliable.  Tables \ref{table_RVs_m1}, \ref{table_RVs_m2}, and  \ref{table_SOFIN_RVs} list the RV values that were used in the analysis discussed below. 

\begin{table}
\begin{center}
\begin{small}
\caption{Fits to RVs: 2000 SOFIN observations\label{table_fits_fotel1}}
\begin{tabular}{lllll}
\hline
\hline
{\bf Param}		&{\bf a) All RVs}  	&{\bf b) No $m_2$ near} 	& {\bf c) $+$16 } 	& {\bf b)+c) }  \\
{}				&				&{\bf conjunctions}		& {\bf km s$^{-1}$}	&         \\
\hline
\hline
$\vv_0$             & $-$1.7(0.2)   &$-$2.5        &6.8(0.4)            & 3.6(0.3)     \\
$e$                    &  0.099         &0.099         &  0.097             &0.124              \\
 $\pm$                 &  (0.002)       & (0.004)      &  (0.003)           & (0.002)       \\
$\omega_{per}$         &  233.8(0.1)    &233.0(0.004)  &  233(0.2)          &233 (0.2)      \\
 $K_1$                 &  121(0.5)      &122(0.8)      &  118(0.7)          &122(0.5)      \\
 $K_2$                 &  188           &188           &  192               &194         \\
$m_1/m_2$              & 0.644          &0.649         &  0.615             &0.627              \\
   $\pm$               & (0.002)        &  (0.005)     &  (0.002)           & (0.003)          \\
$f_1(m)$               &  0.737         &0.748         &  0.682             &0.733        \\
$m_1$                  &   9.7          &9.8           &  10.0              &10.3               \\
$m_2$                  &  6.3           &6.3           &  6.2               & 6.5                \\
$\phi_{conj}$          & 0.084          &0.085         & 0.084              & 0.081             \\
                       & 0.619          &0.623         &0.622               &0.629             \\
\hline
\hline
\end{tabular}
\end{small}
\end{center}
{\bf Notes:} {$\vv_0$ is the systemic velocity, and $K_1$ and $K_2$ the semi-amplitudes of the RV curve, all in units of km s$^{-1}$; $e$ is the eccentricity of the orbit; $\omega_{per}$ is the argument of periastron in units of $deg$;  $f(m_1)$ is the mass function of $m_1$ in units of $M_\odot$; $m_1$, $m_2$ are given in $M_\odot$; $\phi_{conj}$ are the orbital phases of conjunction. The values in parentheses are the uncertainties given by FOTEL.}
\end{table}

\begin{table}
\begin{center}
\begin{normalsize}
\caption{Fits to RV data: 2000 (MTT), 2008, 2013 \label{table_fits_fotel_other}}
\begin{tabular}{llll}
\hline
\hline
{\bf Param}            &{\bf MTT$^b$}    & {\bf 2008}          & {\bf 2013}    \\
\hline
\hline
$\vv_0^a$ km s$^{-1}$ & $-$2(1)          &1.3(2)              & 4.8(1.5)     \\
 $e$                   &  0.100(0.009)   & 0.124(0.008)       & 0.102(0.005)      \\
$\omega_{per}$         &  232(1)         & 258(0.6)           & 277(0.4)      \\
 $K_1$  km s$^{-1}$    &  122(5)         &  122(2)            & 123(1)       \\
 $K_2$  km s$^{-1}$    &  193            &  188               & 192        \\
$m_1/m_2$              &  0.635(0.014)   & 0.648(0.012)        & 0.639(0.005)      \\
$f_1(m)$ $M_\odot$     &  0.749          & 0.733              & 0.762       \\
$m_1$ $M_\odot$        &  10.3           &  9.6               & 10.3              \\
$m_2$ $M_\odot$        &  6.5            &  6.2               &  6.6               \\
$\phi_{conj}$          & 0.087, 0.627    & 0.026, 0.544       & 0.984, 0.476      \\
\hline
\hline
\end{tabular}
\end{normalsize}
\end{center}
{\bf Notes:} {\small $^a$The value of $\vv_0$ in 2008 and 2013 is the average of the result obtained for each of the 4 absorption lines that were measured and the uncertainty is the standard deviation of these measurements. $^b$The fitted data here also exclude most values of $m_1$ near conjunctions which, when plotted on the RV curve, are clearly inconsistent.}
\end{table}

\subsection{Orbital elements}

\begin{figure} 
\includegraphics[width=1.03\linewidth]{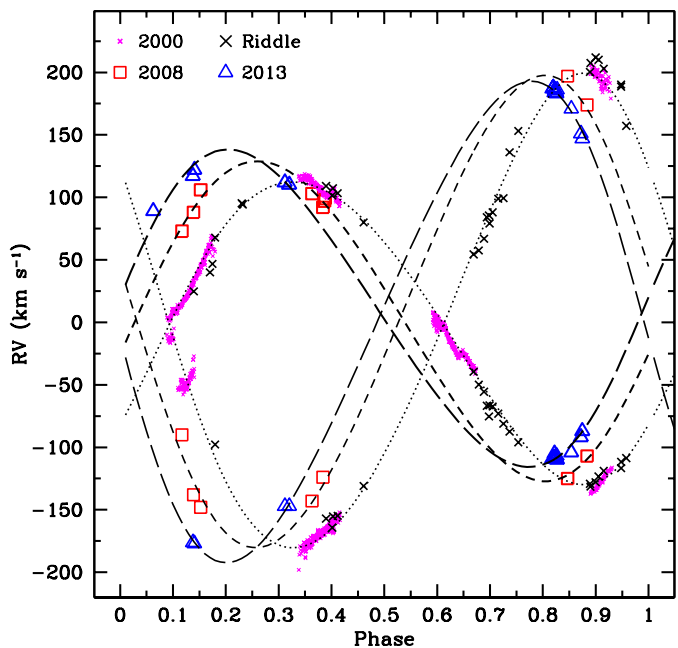}
\caption{Radial velocity measurements of the Si III 4552 \AA\ line from the following data sets: SOFIN 2000 (small overlapping crosses), {\rm MTT} 2000 (large crosses), and ESPaDOnS 2008 (squares) and 2013 (triangles).  The curves correspond to the best fits obtained with FOTEL.
\label{rv_sbcm_m1m2}}
\end{figure}

        The orbital elements were obtained from fits  to the RV curves using the program FOTEL \citep{2004PAICz..92....1H}.  The RVs of $m_1$ and $m_2$ from the SOFIN data set were fit simultaneously, holding only  the orbital  period and the initial epoch constant and yielded the results listed in Column 2 of Table \ref{table_fits_fotel1}.  The RVs of the four lines measured in the 2008 and  2013 data sets for  $m_1$ and $m_2$ were also fit simultaneously for each epoch, and the result listed in Columns 3 and 4 of Table \ref{table_fits_fotel_other}.  Finally, the same procedure was applied to the Si III RVs given by \citet{2000PhDT.......148R}, referred to hereafter as the {\rm MTT}-data set. The latter, however, yielded  value of $e$ and $\omega_{per}$ that differed significantly from those derived from the SOFIN data set (the {\rm MTT} data gave $e$=0.05$\pm$0.02, $\omega_{per}$=223$\pm$2).  Upon further inspection, it is evident that several RV values in the {\rm MTT}-data set, particularly those that were deconvolved at orbital phases around conjunctions,  are in error.  Once these are eliminated, results that are fully consistent with those of the SOFIN data are obtained, and these are listed in Column 2 of Table \ref{table_fits_fotel_other}.  The observed RVs and the best fit curves are plotted in Fig. \ref{rv_sbcm_m1m2}.

Note that the 2013 data set lacks spectra at orbital phases near conjunctions.  Similarly, the data points that were eliminated in the {\rm MTT}-data correspond to this same phase range, and we made not attempt to measure $m_2$ in the 2008 spectra in which it was severly blended with $m_1$.  In order to evaluate the effect of neglecting the RVs of $m_2$ around conjunctions and the possible effect of systematic shifts in the $m_1$ RVs at these phases, we ran the following tests using the SOFIN data set with the corresponding results, summarized in Table \ref{table_fits_fotel1}:  1) We eliminated all the RVs of $m_2$ near conjunctions, which had no effect on the fit to the RV curve (Column 3).  2) We introduced an artificial $+$16 km s$^{-1}$ shift to the RVs of $m_1$ around conjunctions to take into account possible systematic shifts due to the tidal flows, which led to a negligible decrease of $e$ (Column 4). 3) We eliminated the RVs of $m_2$ near conjunctions and introduced the $+$16 km s$^{-1}$ shift, which led to $e$=0.124 (Column 5).  The longitude of periastron, $\omega_{per}$ was unaffected.

The basic conclusion is that the fits to the 4 RV curves yield $e$=0.108$\pm$0.014, where the uncertainty corresponds to the standard deviation of the four fitted values about their mean. This uncertainty is 2 to 7 times larger than that given by the formal error in the FOTEL fit and we believe that it is a more realistic value, particularly in light of the dependence that we find below on the epoch of observations. 

The other fit parameter that we consider relevant to mention here is $\vv_0$, the sistemic velocity. There is a systematic trend over time starting with $\vv_0$=$-2$ km s$^{-1}$ in 2000 and ending with $\vv_0$=$+$4.8 km s$^{-1}$  in 2015. It is not clear at this time whether this range, which is more than 3 times the quoted uncertainty in the fit, is significant.  From the tests we performed on the SOFIN data set, the trend can be made to vanish by introducing artificial shifts to the $m_1$ RVs around conjunctions, so it may simply be a consequence of the line-profile variability viewed in different geometrical configurations over time.  However, we cannot at this time exclude the possibility of a variable systemic velocity which would suggest the presence of a third object in the system.

\subsection{Apsidal period}

        The values of the longitude of periastron determined for each of our observation epochs were combined with values obtained from observations by \citet{1910PAllO...1...65B}, \citet{1934ApJ....80..365S} and \citet{Struve:1958kz} listed in \citep{1972MNRAS.156..165S}, listed in Table \ref{table_apsidal_period}.\footnote{Note that the value quoted for Herbison-Evans et al. (1971) in this table corresponds to that listed in their Table II and not to the one they give as a final value, $\omega_{per}$=138, since the latter had been fixed for their analysis and had no uncertainties associated.} A linear fit to $\omega_{per}$ {\it vs.} Julian Date (JD) yields

\begin{equation}
\omega_{per}(t)=\dot{\omega}_{per}t-197^\circ\pm7^\circ
\end{equation}

\noindent where $t=T_{JD}-2400000$, with $T_{JD}$ the date observation in Julian Days, $\omega_0$=197$^\circ\pm7^\circ$ is the y-intercept of the fit, and $\dot{\omega}_{per}=\frac{d \omega_{per}}{dt}=$(8.36$\pm$0.13)$\times$10$^{-3}$ deg d$^{-1}$ is the rate of change of the longitude of periastron. Setting the initial epoch from \citet{HerbisonEvans:1971th} to T$_0=JD2440678.09$, $\omega_{per}(T_0)=142.7^\circ$.

The period of rotation of the line of apsides is  $U=\frac{360}{\dot{\omega}_{per}\times 365.25}=117.9\pm 1.8$ {\rm yrs}.  This value is very close to that arrived at by \citet{2015AAS...22534520A}, and lies within the uncertainties of the value quoted by \citet{HerbisonEvans:1971th}.

\begin{figure} 
\begin{center}
\includegraphics[width=1.02\linewidth]{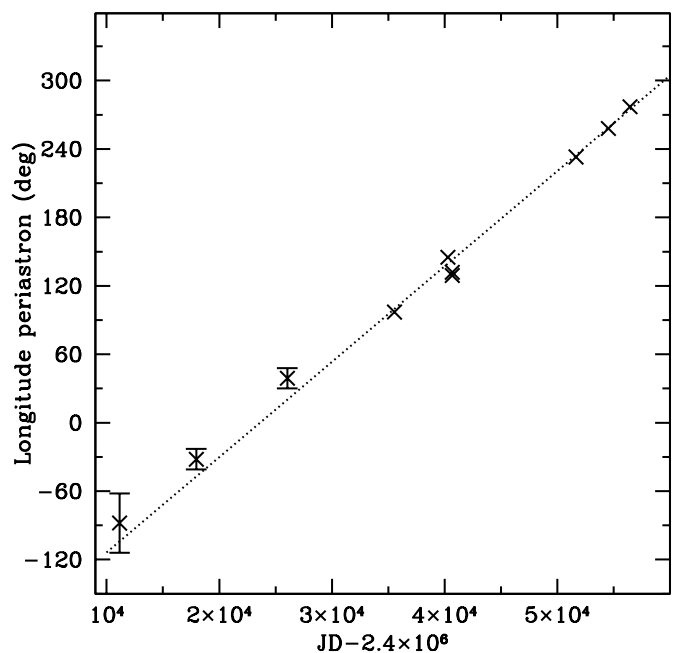}
\caption{\label{fit_omega} Values of the longitude of periastron ($\omega_{per}$) as a function of epoch of observation (see Table~\ref{table_apsidal_period}).  The line is a linear fit to the data. The error bars for the years 1956, 1969 and 1970 are the size of the symbols and they are smaller than the symbols for 2000, 2008 and 2013.}
\end{center}
\end{figure}

\begin{table}
\begin{center}
\begin{normalsize}
\caption{Longitude of periastron\label{table_apsidal_period}}
\begin{tabular}{llrrrrrr}
\hline
\hline
{\bf Epoch}     & {\bf JD}              & {\bf $\omega_{per}$}  & {\bf Reference}            \\
{$Year$}     & {$-2400000$}       & { $deg$}  & {}            \\

\hline
1889                    &  11158                        &   272 $\pm$    26     &  estimated$^3$   \\
1908                    &  17955                        &   328 $\pm$     9     & Allegheny$^2$  \\
1934                    &  26041                        &    39 $\pm$     9     & Yerkes$^2$   \\
1956                    &  35563                        &    97 $\pm$     8     &  $^4$  \\
1969                    &  40284                        &   145 $\pm$     7     & $^5$  \\
1970                    &  40678                        &   132 $\pm$     8     & HE71 \\
1970                    &  40690                        &   129 $\pm$     7     & Duke (1974) \\
2000                    &  51654                        &   233 $\pm$     0.5   & This paper  \\
2008                    &  54547                        &   258 $\pm$     0.5   & This paper  \\
2013                    &  56451                        &   277 $\pm$     0.5   & This paper  \\
\hline
\hline
\end{tabular}
\end{normalsize}
\end{center}
{\bf Notes:} {\small  
$^2$redetermined by \citet{1972MNRAS.156..165S}, $^3$\citet{Luyten:1935ec}, $^4$\citet{1972MNRAS.156..165S}, $^5$computed in \citet{1972MNRAS.156..165S}.}
\end{table}

\section{Theoretical line profile variability} 

In general, spectral line profile variability is produced by perturbations on a star's surface which cause patches of photosphere to move at speeds that are different from that of a rigid body rotator or have different than average effective temperatures and gravities. The source of the perturbations can be intrinsic (pulsations, magnetic fields, convective cells reaching the surface, patches in which radiation pressure is more effective thus causing a localized stellar wind) or caused by the companion (irradiation and tidal forces).

The fact that some of the line-profile variations in Spica have periodicities that are sub-multiples of the orbital period points to binary-induced effects.  Since effects of irradiation by $m_2$ on $m_1$ in Spica have been shown to be negligible \citep{Palate:2013ew}, it is the tidal forces which come into focus. \citet{Smith:1985bd,Smith:1985dg}  assumed that the variability could be modeled as non-radial pulsations and proceeded to fit the data with spherical harmonics to derive periodicities for the migrating ``bumps''. However, he concluded that features such as the ``red spike'' are actually associated with the horizontal component of the surface velocity field which is primarily induced by the tidal interactions.

Our approach has been to make no assumptions regarding the origin of the surface perturbations.  Rather, we constructed a dynamical model that takes into account the forces that are present on a rotating binary star's surface throughout its orbital cycle. As we have previously shown, many of the features of Spica's variability can be described with this model.  The tidal perturbations depend strongly on the ratio of stellar radius to orbital separation, $R_1/a$, so if the ``red spike'' is indeed  caused by the horizontal velocity field it could provide a diagnostic for $R_1$.  

The {\it Tidal Interaction with Dissipation of Energy by Shear (TIDES)} code \citep{1999RMxAA..35..157M, Moreno:2005cq, Moreno:2011jq} computes from first principles the photospheric line profiles produced by a tidally-perturbed surface layer of depth $dR$ by solving the equations of motion of volume elements that make up an outer shell that covers the rigidly rotating inner body of the star.  {\it TIDES} does not fit line profiles. It is a time-marching algorithm that, once the input parameters are specified, computes the perturbations on the 
stellar surface, projects the velocity field along the line-of-sight to the observer, shifts local line profiles at each surface element accordingly, and produces the integrated line profiles at each orbital phase over the orbital cycle.  Thus, the output of a {\it TIDES} calculation consists of a set of line profiles as a function of orbital phase.

For the model calculations, the values listed in Column 3 of Table 1 were used for the input parameters $m_1$, $m_2$, $P$, $i$, $\vv_{rot1}$, and the value $e$=0.100.  Using $e$=0.124 does not modify the general conclusions of this section.   The layer depth is $dR$=0.07 $R_1$. The examples shown below were run with different values of $R_1$ and it is important to keep in mind that the observed rotation velocity of $m_1$  combined with ($R_1$) yields the asynchronicity parameter ($\beta_0$).  With $R_1$=6.84 -- 8.4 R$_\odot$ and the fixed value for v$_{rot}$, these radii correspond to asynchronous rotation rates at periastron $\beta_0$=$\omega/\Omega_0$=1.52 -- 1.24, respectively, where $\omega$ is the rotation angular velocity and $\Omega_0$ is the orbital angular velocity at periastron.  

Once the input parameters are specified, there is no further intervention by the user in the computation.  The calculation proceeds with the time steps that are required by the integration of the equations of motion,  and produces output only at predetermined times.  The product of each run is a set of line profiles for each orbital phase that was pre-specified. Thus, {\it TIDES} does not fit line profiles, instead it produces a sequence of profiles, corresponding to the sequence of pre-specified orbital phases, which can then be compared with an analogous sequence of observational data.   The  model is currently limited due to its  one-layer treatment of the problem.  The consequence of this limitation is that it ignores the non-radial pulsations due to the normal  modes of the star, hence neglecting the interplay between these modes and the tidal flows.  

\begin{figure} 
\begin{center}
\includegraphics[width=0.32\linewidth]{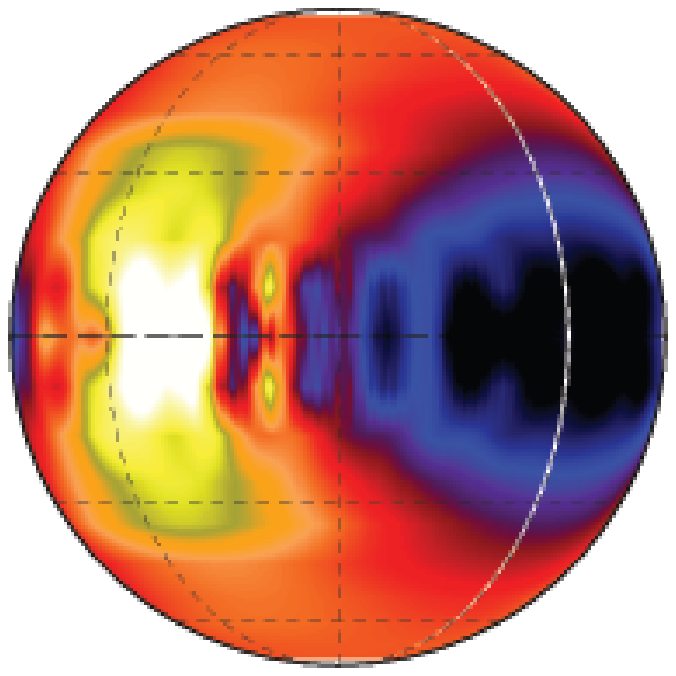}
\includegraphics[width=0.32\linewidth]{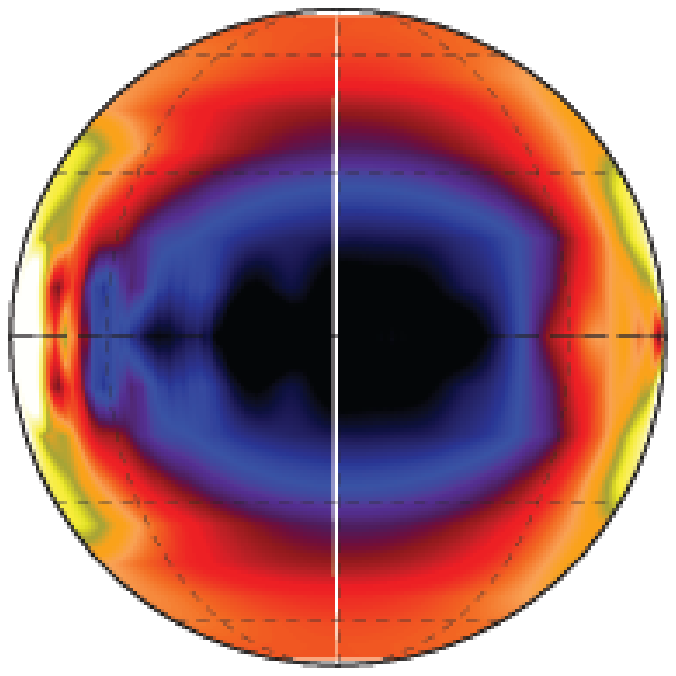}
\includegraphics[width=0.32\linewidth]{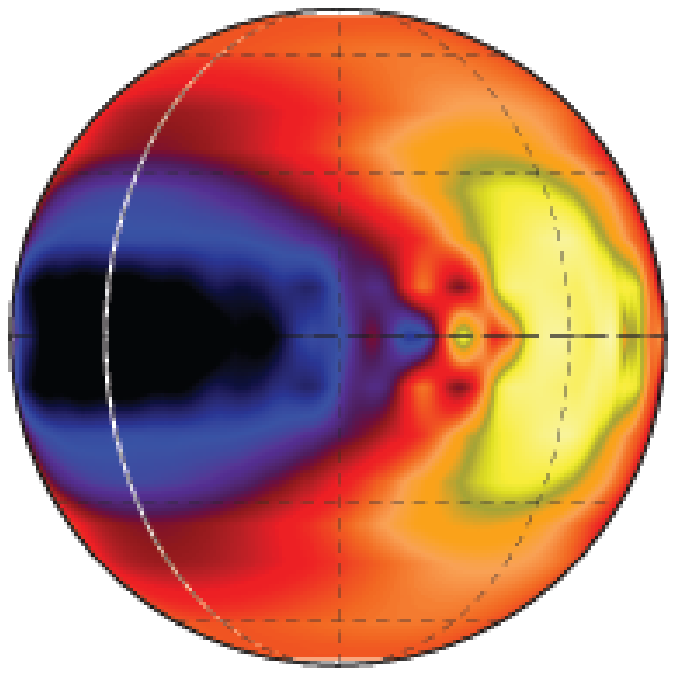}
\includegraphics[width=0.99\linewidth]{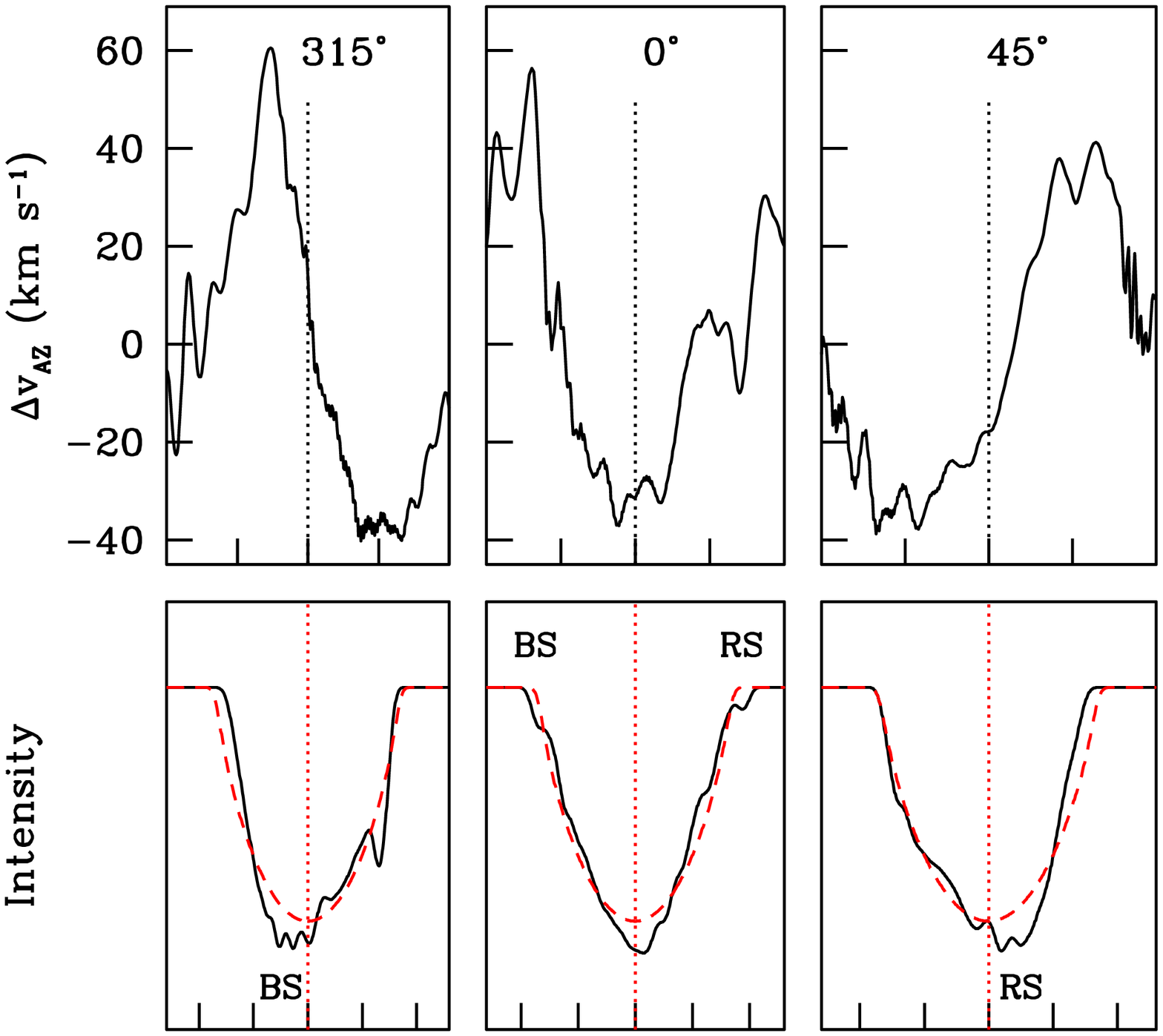}
\caption{\label{velocity_field} {\it Top:} The residual surface horizontal velocity field, $\Delta \vv_{AZ}$, on the hemisphere that would be seen by an observer around the time of conjunction when $m_2$ is in front of $m_1$ for the $R_1$=7.6 $R_\odot$, $e$=0.1 calculation and for the epoch 2000, when $\omega_{per}$=233$^\circ$.  From right to left: half a day before conjunction ($\phi$=0.494), at conjunction ($\phi$=0.619) and half a day after ($\phi$=0.750).  The maps are centered, respectively, on longitudes $\varphi$=45$^\circ$, 0$^\circ$, 315$^\circ$.  The sub-binary longitude is shown by the white dotted line that crosses the dark patch. The sense of rotation is such that the left limb is approaching the observer; i.e., rotation is counterclockwise when viewed from the north pole, and so is the orbital motion. Darkest patches correspond to tidal flows with $\Delta \vv_{AZ}<0$  and lightest patches to $\Delta \vv_{AZ}>0$.  {\it Middle:}  $\Delta \vv_{AZ}$ along the equator corresponding to the maps above. The ordinate is in units of $km~s^{-1}$. The tick marks in the abscissa are set at 45$^\circ$ intervals with the central longitude indicated with the dotted line. {\it Bottom:} The corresponding line profiles compared with the non-perturbed line profiles (dash). The dotted line indicates the zero velocity and the ticks are separated by 100 $km~s^{-1}$. {\it BS} and {\it RS} indicate the blue and red spikes.
}
\end{center}
\end{figure}

\subsection{The velocity field that causes the variability}

The results of our computations show that the horizontal velocity perturbations are generally an order of magnitude larger than those in the radial direction, so we focus here on the residual horizontal velocity field;  that is, $\Delta \vv_{AZ}=\vv_\varphi(\varphi,\theta)-\vv_{rot}$, where $\vv_\varphi(\varphi,\theta)$ is the horizontal velocity of a volume element located at polar angle $\theta$ and azimuth angle $\varphi$, measured with respect to the sub-binary longitude in the direction of stellar rotation, and $\vv_{rot}$ is the constant rotation rate of the underlying, assumed rigid body of the star.  The maps in Fig. \ref{velocity_field} display the color-coded values of $\Delta \vv_{AZ}$ on the hemisphere that would be seen by an observer at the time of conjunction when $m_2$ is in front of $m_1$,  just before this time and just after.  For the calculation with $\omega_{per}$=233$^\circ$, these times correspond to orbital phases, 0.62, 0.50, and 0.76, respectively.

These maps show that the stellar surface is divided up into large patches of faster- and slower-than rotation regions and these lead to structural changes in the overall line profiles as velocity perturbations rotate in and out of the line of sight.  The values of $\Delta \vv_{AZ}$ are considerable, as illustrated in the middle plots in Fig. \ref{velocity_field}, where its value along the equator is plotted for the same hemisphere that is illustrated in the map.  

The bottom set of plots displays the line-profile in the observer's frame of reference, for the same orbital phases and for the case of an orbital inclination $i$=66$^\circ$.  The perturbed profiles are compared with the profile they would have were no perturbations present, and the presence of blue/red spikes are indicated.  Prior to conjunction, there is a prominent red spike and the red line wing has a steeper slope than the blue wing.  After conjunction, the spike is now on the blue side of line center, and the red spike has now moved out to $\sim$125 km s$^{-1}$ with a significantly narrower shape.  At conjunction, both a blue and a red spike are seen forming near the continuum level.

From a qualitative standpoint, the comparison with the observational data (Fig. \ref{trends_block4}) is positive.  However, the strength and location of the BS and RS do not coincide.  Specifically, these spikes are significantly stronger in the observations than shown in Fig. \ref{velocity_field}, and their locations differ.  These discrepancies can be attributed to the limitations of the one-layer model used in our calculations and also to the fact that the model does not take into account possible temperature variations over the surface which, in addition to altering the strength of the local line profile could introduce a time-dependence into the limb-darkening coefficient. It is also important to note that our model assumes that the $m_1$ and $m_2$ axes of stellar rotation are perpendicular to the orbital plane.  Departures from this condition would lead to a different surface velocity field. Some of these issues will be addressed in the future.

\begin{figure} 
\includegraphics[width=1.03\linewidth]{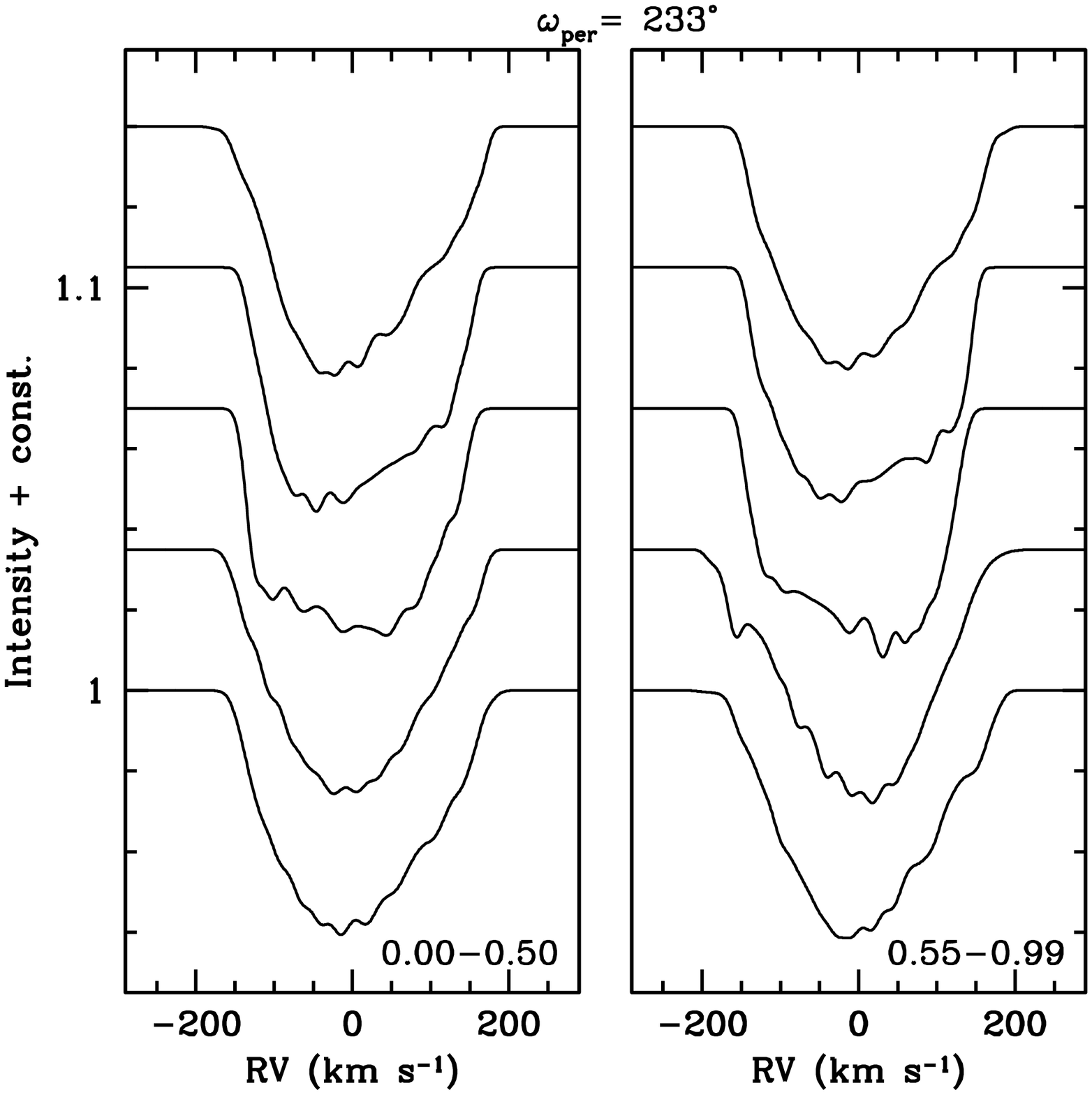}
\includegraphics[width=1.03\linewidth]{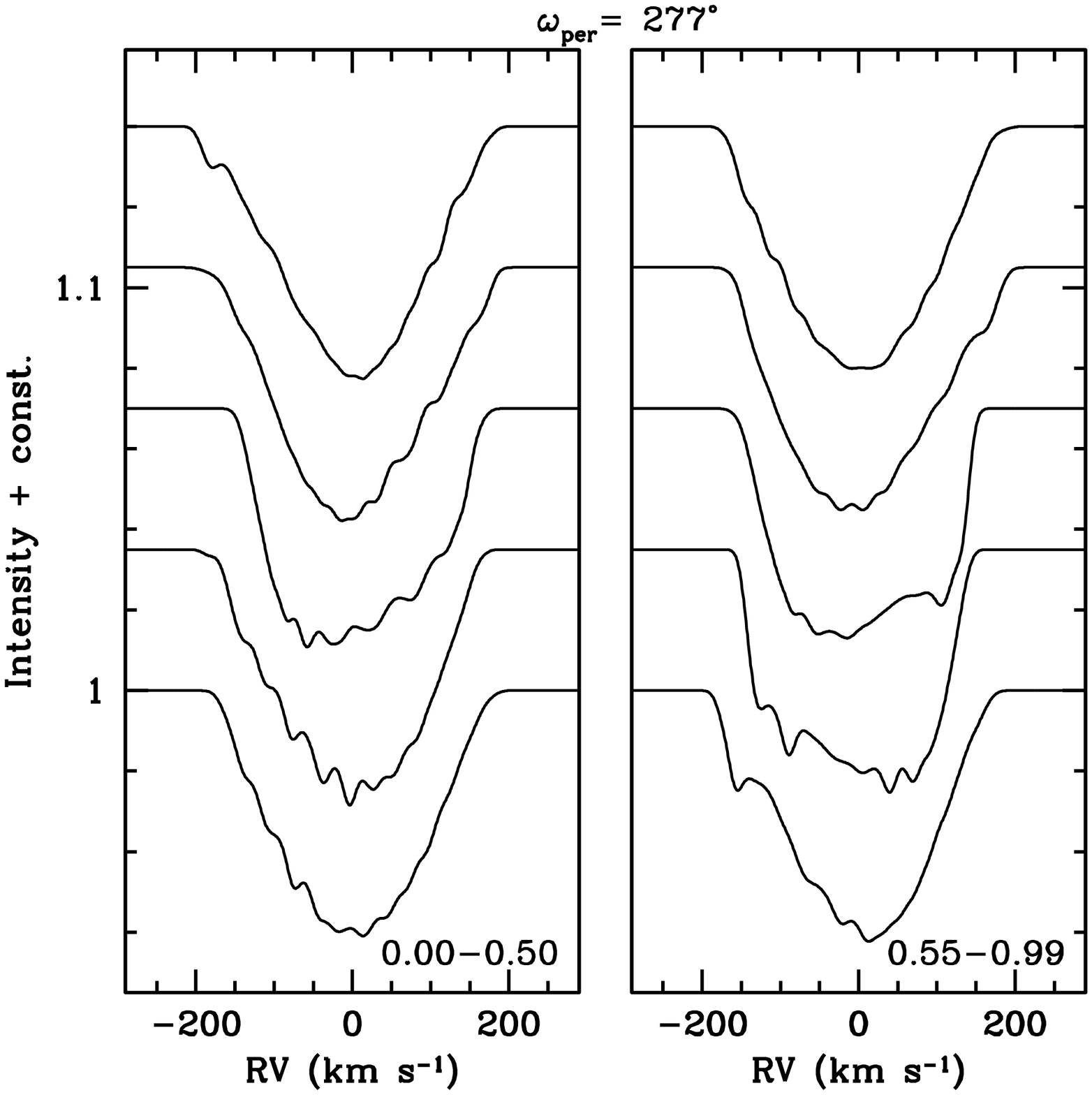}
\caption{Model line profiles for a primary star with $R_1$=7.6 $R_\odot$ and for $\omega_{per}$=233$^\circ$, illustrating the transition in shape from parabolic-like to  ``boxy'' (i.e., having a flat-bottom and steep wings), the bottom of which then slopes upward towards the blue, until the parabolic-like shape is recovered.  The sequence from periastron to apastron is shown on each of the left panels and from apastron to periastron on the right. Time increases upward.  Note the symmetry, caused the by geometry of the tidal flows and the orbital motion. The two panels on the left correspond to $\omega_{per}$=233$^\circ$ and the two on the right to $\omega_{per}$=277$^\circ$.
\label{model_omegas}}
\end{figure}

\subsection{From parabolic to ``boxy'' and back}

Figure \ref{model_omegas} illustrates the theretical line profile variability for the orbital phase intervals going from periastron to apastron (left) and from apastron to periastron (right). The first point to note is that the same trend in the line-profile shape is evident in both trajectories. The profile starts with a parabolic-like shape, develops blue and red spikes and several ``bumps'', attains a flat-bottom (``boxy'') shape, the bottom of which then slopes upward towards the blue, until the profile recovers the parabolic-like shape.   The direction of the slope at the core of the ``boxy'' line profiles undergoes the transition from negative to positive over the orbital phase intervals 0.30-0.42 and 0.80-0.92.   This is the same behavior that we show in Section 3 is present in the SOFIN data set (Fig. \ref{trends_block2}).  The fact that a similar cycle occurs twice during an orbital cycle is due to the geometrical configuration of the tidal flow structure on the stellar surface.

It is important to note is that the geometry of the tidal flow velocity field is relatively fixed in the frame of reference that rotates with the binary companion; i.e., the primary ``bulge'' points towards the companion. However, the projection of this field along the line-of-sight to the observer changes as a function of orbital phase.  This causes the orbital phase-locked line profile variability that we observe.

Over long timescales, the rotation of the line of apsides introduces changes in the geometry and projected velocity field of the stellar surface as viewed from Earth at specific orbital phases, thus making line profiles that are observed at the same orbital phase to appear different from those at the same orbital phase at other  epochs.  Hence, in Spica, the orbital phase-locked variations are locked more with respect to the phase of conjunctions, rather than phase with respect to periastron.  This is not necessarily the case in other binary systems having larger eccentricities.

\subsection{The eccentricity problem}

\begin{figure} 
\includegraphics[width=1.05\linewidth]{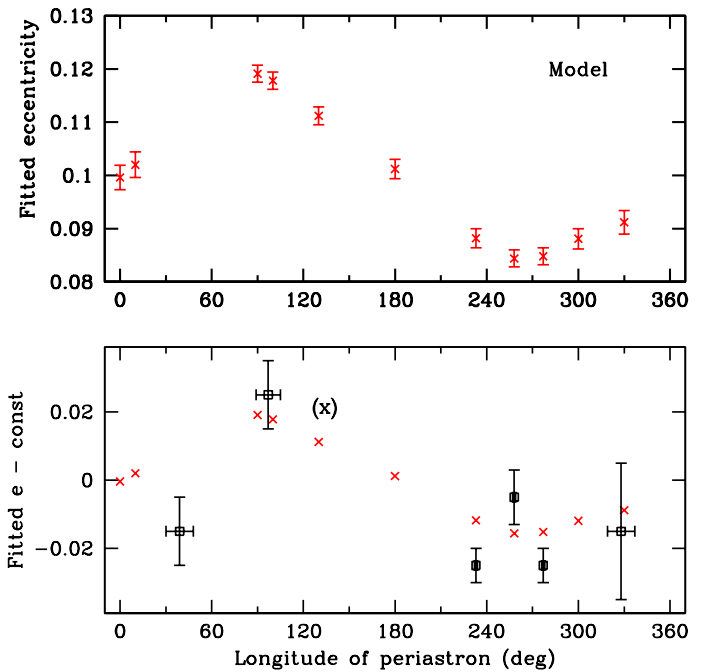}
\caption{{\it Top:} Eccentricity fitted by SBCM \citep{1974PASP...86..455M} using the RVs measured on the theoretical, tidally-perturbed line profiles of $m_1$ as a function of the longitude of periastron.  The calculations were all performed with $e$=0.1.  The error bars are the uncertainties given by the RV curve fitting routine. {\it Bottom:} The crosses are the fitted eccentricities shown in the top panel  after subtracting the actual value $e$=0.1.  The squares are the historic $e$ values quoted by \citet{1972MNRAS.156..165S} and the values we have derived in this paper, after subtracting the mid-point value $e$=0.125.  The value in parenthesis is the one given in \citet{HerbisonEvans:1971th}. The error bars represent the corresponding quoted uncertainties. 
\label{eccentricity_omega}}
\end{figure}

A series of  {\it TIDES} models were run with $R_1$=7.6 $R_\odot$ and $e$=0.100, and holding all the parameters fixed except for $\omega_{per}$, which was varied from 0$^\circ$ to 360$^\circ$. The centroids of the theoretical line profiles  were measured using the same method as applied to the observations.  The resulting RVs were then fed into FOTEL in order to retrieve the orbital elements.  The results, shown in Fig. \ref{eccentricity_omega}, show that the  fitted value of $e$ makes an excursion of approximately $\pm$0.02 around its true value.  Thus, the fitted value of $e$ {\em depends on the value of $\omega_{per}$}.  This means that in  real binary systems with relatively short apsidal periods, one might derive a value of $e$ that depends on the epoch of observation.

The range of $e$ values obtained for {\it Spica} by observers dating back to 1908 is 0.10 -- 0.15 (excluding the value of Riddle (2000) which we showed above is the same as  our year 2000 value when the deconvolved data points are eliminated). Subtracting the mid-point of this range (0.125) and plotting as a function of $\omega_{per}$  displays a trend that is similar to that of our model profiles.  From this we conclude that the most likely value of the {\em true} eccentricity of Spica is $e$=0.125, listed as the second value of $e$ in Column 3 of Table 1.

\subsection{Changes in profiles as a function of $R_1$}

\begin{figure} 
\includegraphics[width=1.03\linewidth]{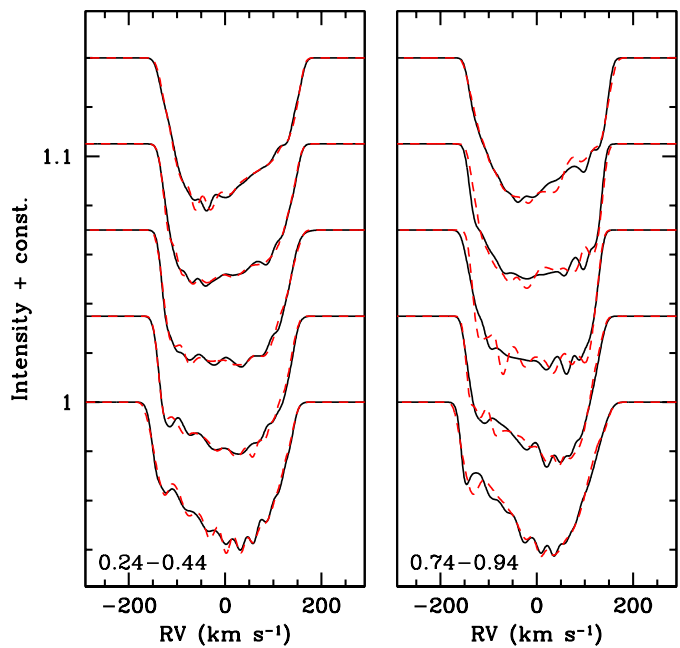}
\caption{Theoretical line profiles for $m_1$ with $R_1$=7.6 $R_\odot$ and 8.4 $R_\odot$ (dash) in the orbital phase intervals 0.24-0.44 and 0.74-0.94 for $\omega_{per}$=233$^\circ$.  The orbital phases run from bottom to top, as indicated. Note that the most significant differences between the profiles of different radii is in the ``bumps''  as the star approaches periastron (right).
\label{model_compare_radii}}
\end{figure}

In order to address the question of whether the tidally-induced line profile variability can be used to constrain the stellar radius, $R_1$ of Spica's primary, we computed a model for each of of the following radii, 6.84, 7.6 and 8.4 $R_\odot$.  The results of this exploratory analysis is that, holding all parameters constant other than $R_1$, the value of the radius affects the time of appearance and location of the red spike and the number, strength and location of the other ``bumps''.   This is illustrated in Fig. \ref{model_compare_radii}, which also shows that the large-scale structure of the line profiles is not strongly affected by the different radii.   

As an example of the type of test that may be performed once an n-layer calculation is available, we compare in Fig. \ref{compare_spikes_peri1.eps} two of the observed line profiles from the SOFIN data set with the corresponding model line profiles from the theoretical set described above. This comparison, taken at face value, would indicate that the model computed with $R_1$=7.6 $R_\odot$ produces a red spike whose location agrees better with that of the observations than that of the other two models.

\begin{figure} 
\includegraphics[width=1.03\linewidth]{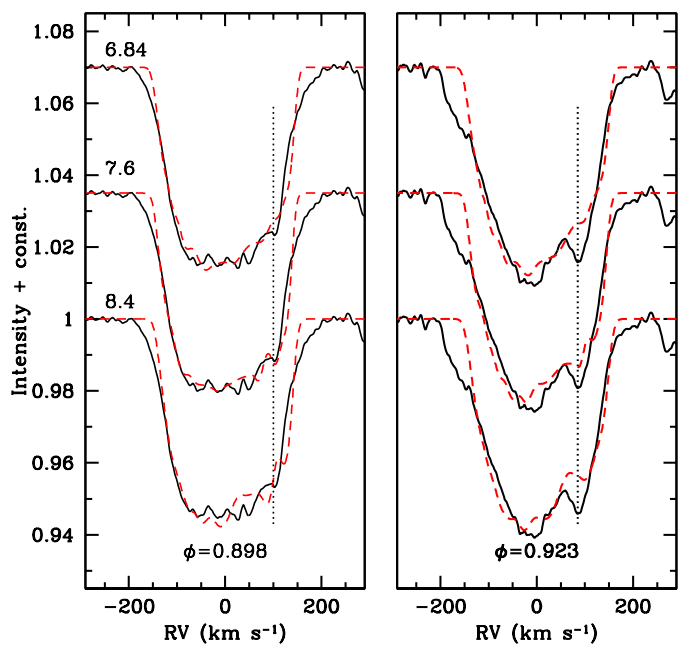}
\caption{Observed SOFIN spectra at orbital phase 0.898 (left) and 0.923 (right) compared to line profiles computed with {\it TIDES} (dash) using the range in values of $R_1$=6.84, 7.6 and 8.4 R$_\odot$. The dotted line indicates the location of the ``red spike'' in the observations: 100 km/s (left) and 85 km/s (right).  The ``red spike'' in the $R_1$=8.4 R$_\odot$ is clearly discrepant with the data, while the models with the smaller radii seem to provide a better agreement. 
\label{spike_peri1}}
\end{figure}

\section{The internal structure constant}

 \citet{2002A&A...388..518C}  provide an expression for calculating the value of $k_{2,obs}$, the observational interior structure constant:

\begin{equation}
k_{2,obs}= \frac{1}{c_{21}+c_{22}} \frac{P}{U}
\end{equation}

\noindent where $c_{2i}=f_i(\beta_i,q,e) (R_i/a)^5$, with $i$-1,2 indicating primary and secondary, respectively,  $\beta_i$ the ratio of stellar rotation angular velocity to mean orbital angular velocity, $q=m_1/m_2$, $a$ the orbital semi-major axis, and $f_i(\beta_i,q,e)$ functions of the variables enclosed in the parentheses.

We used this expression to determine  $k_{2,obs}$ using the parameters we have refined in this paper and  $R_1/R_\odot \in [6.84,8.4]$. The smallest value in this range was derived by \citet{Palate:2013ew} from stellar atmosphere fits to Spica's spectrum, and the largest value comes from the  interferometric+spectroscopic results of HE71 who found $R_1$=8.1$\pm$0.5 R$_\odot$.  An intermediate value of 7.6 $\pm$0.2 $R_\odot$ was obtained from the  ellipsoidal light variations of \citet{Sterken:1986tp}.\footnote{Corresponding to the polar radius, and the authors note that in order to produce the observed light variations, the equatorial radius would have to be $\sim$4\% larger than the polar radius.  Thus, they note that the interferometrically determined radius (presumably a mean value over the entire star) would be $\sim$2\% larger than the polar radius.}

The values of $k_{2,obs}$ that result for  $e$=0.100, $U$=117.9 yr, $m_1$=10 $M_\odot$, $m_2$=6.3 $M_\odot$, $\vv_{rot,1}$=176 $km s^{-1}$, $v_{rot,2}$=70 $km s^{-1}$, are listed in Table ~\ref{table_k2obs}.  The derived range is $k_{2,obs} \in [-2.66,-2.28]$.  Table~\ref{table_k2obs} also lists results for a different asynchronicity parameters; that is, instead of using $<\beta> =(\frac{\omega_i}{\Omega_k})$ as defined in the equation above, we use the asynchronicity at periastron, $\beta_0=(\frac{\omega_i}{\Omega_{per}})$. This does not significantly change these numbers, nor does a value of $e$=0.125. 

  The theoretical value for a star with $m_1$=10 $M_\odot$, and $log(g)^{pole}$ and $T_{eff}$ as given by \citep{Palate:2013ew} can be found in Table 31 of \citet{1991A&A...244..319C}: $log(k_{2,theory})=-2.27$. A slightly larger value, $log(k_{2,theory})\sim -2.4$, is given by  Claret \& Gim\'enez (1993, Fig. 9).\footnote{Spica is denoted by a cross in this figure, and since $log(k_{2,obs})\sim$-2.6, one needs only to find the value of the abscissa corresponding to this point.} Thus,  Spica's observational internal structure is consistent with theory for $R_1=6.8-7.4 R_\odot$.

\begin{table}
\begin{center}
\begin{large}
\caption{Internal structure constant\label{table_k2obs}}
\begin{tabular}{llrrrrrr}
\hline
\hline
{\bf R$_1$/R$_\odot$}  & {\bf $\beta_0$} & {\bf $<\beta>$} & {\bf log($k_{2,obs}$)$^{\beta_0}$}  & {\bf log($k_{2,obs}$)$^{<\beta>}$}     \\
\hline
8.4          &  1.37           & 1.44               &   -2.65             & -2.66                          \\
8.2          &  1.40           & 1.48               &   -2.60             & -2.61                          \\
8.0          &  1.44           & 1.51               &   -2.55             & -2.57                          \\
7.6          &  1.51           & 1.59               &   -2.46             & -2.47                          \\
6.84         &  1.68           & 1.77               &   -2.27             & -2.28                          \\
\hline
\hline
\end{tabular}
\end{large}
\end{center}
{\bf Notes:} {\small {$\beta_0$ is the ratio of rotation to orbital angular velocity at periastron; $<\beta >$ is the average value over the orbital cycle; $k_{2,obs}^{\beta_0}$ and $k_{2,obs}^{<\beta>}$ are the values of the observational internal structure constant computed with Eq. (4) of Claret \& Willems (2002) using $\beta_0$ and $<\beta >$, respectively.}
}
\end{table}

\section{Conclusions}

In this paper we report the analysis of  high spectral resolution and high signal-to-noise data sets of the Spica binary system obtained in 2000, 2008 and 2013. These are used to analyze the line profile variability and derive a refined set of orbital elements. We use theoretical sequences of line profiles computed with the {\it TIDES} code to study the manner in which the radius of the primary star ($R_1$) and the orientation of the orbit with respect to the observer ($\omega_{per}$) affect the time-dependent shape of the line profiles.  

We find that when the radial velocity curve obtained from the tidally-perturbed line profiles is fitted to find orbital elements, the derived eccentricity $e$ depends on the value of $\omega_{per}$, the longitude of periastron.  The variation predicted by our model is consistent with the trend in  $e$  derived from historical data sets combined with our current results.  This leads to the conclusion that the most likely value for Spica's true eccentricity is $e$=0.125, with excursions of $\pm$0.016 due to the epoch-dependent value of $\omega_{per}$.

Our derived values of $\omega_{per}$ are combined with those reported previously to refine the period of precesion of the line of apsides $U$=117.9$\pm$1.8 {\rm yrs}.

The theoretical line profiles show that the general properties of the observed line profile variability can be understood in terms of the tidal flows.  In particular, the transition from a parabolic-like to a ``boxy''-shape and back which occurs twice during the orbital cycle, and the appearance of the blue and red spikes around conjunctions.  However, the smaller-scale structures (``bumps'') may be more related to hydrodynamical effects not captured by the code and non-radial pulsations.

The discrepancy between the observational and theoretical internal structure constant disappears when the stellar radius, $R_1=6.84 R_\odot$,  that was derived by Palate et al. (2013) is adopted.  We note, however, that there are still unresolved issues in the Spica system. These concern the value of $\vv ~sini$ that is deduced from the observations and the orbital inclination $i$. As shown in this paper, sistematic variations (on the order of 15 $km~s^{-1}$) are observed in the half-width of the lines  when comparing the profiles around conjunctions and elongations.  In addition, the peculiar shape of the line profiles precludes a straightforward line-profile fitting technique to derive the projected rotation velocity. Because small differences in $\vv ~sini$  can lead to significant differences in the tidal-flow structure, models of the line-profile variability are still quite uncertain.  

The value of $i$ that we have adopted was obtained from the interferometric observations of \citet{HerbisonEvans:1971th} which, as pointed out by \citet{2007IAUS..240..271A}, neglected the non-constant brightness across the stellar disks of the stars. Preliminary analyses of more recent interferometric observations suggest that $i$ could be significantly smaller \citep{2015AAS...22534520A}.  Clearly, a smaller $i$-value has a large impact on all the derived fundamental parameters.  

These issues highlight the non-linearity of the problems in question and the challenges involved in solving 
the internal structure problem even in a system as close and bright as Spica.

\vskip1cm
\noindent {\bf Acknowledgements.} GK thanks Petr Hadrava for providing the FOTEL code, and R. Gamen for additional guidance on its use; and  Doug Gies for providing the SBCM code. We thank the anonymous referee for very helpful comments and suggestions.  GK acknowledges financial support from CONACYT grant 129343 and UNAM/PAPIIT grant IN 105313, thanks Doug Gies for a useful discussion, and thanks Ulises Amaya and Francisco Ruiz for computer  support. DMH and SVB acknowledge support from the InnoPol grant: SAW-2011-KIS-7 from Leibniz Association, Germany, and by the European Research Council Advanced Grant HotMol (ERC-2011-AdG 291659).  SVB acknowledges the support from the NASA Astrobiology Institute and the Institute for Astronomy, University of Hawaii. This program was partially supported by the Air Force Research Labs (AFRL) through salary support for DMH until 10/2015. This work is partially based on observations made with the Nordic Optical Telescope, operated by the Nordic Optical Telescope Scientific Association at the Observatorio del Roque de los Muchachos, La Palma, Spain of the Instituto de Astrofisica de Canaries.  This work made use of the Dave Fanning and Markwardt IDL libraries. 

\bibliographystyle{aa} 
\bibliography{paper}

\Online

\begin{appendix}

\section{Supplementary figures and tables}

This appendix contains additional figures and the tables in which the measured radial velocites for all our observations are listed.  Figure and table captions are self-explanatory.

\begin{figure} 
\includegraphics[width=1.02\linewidth]{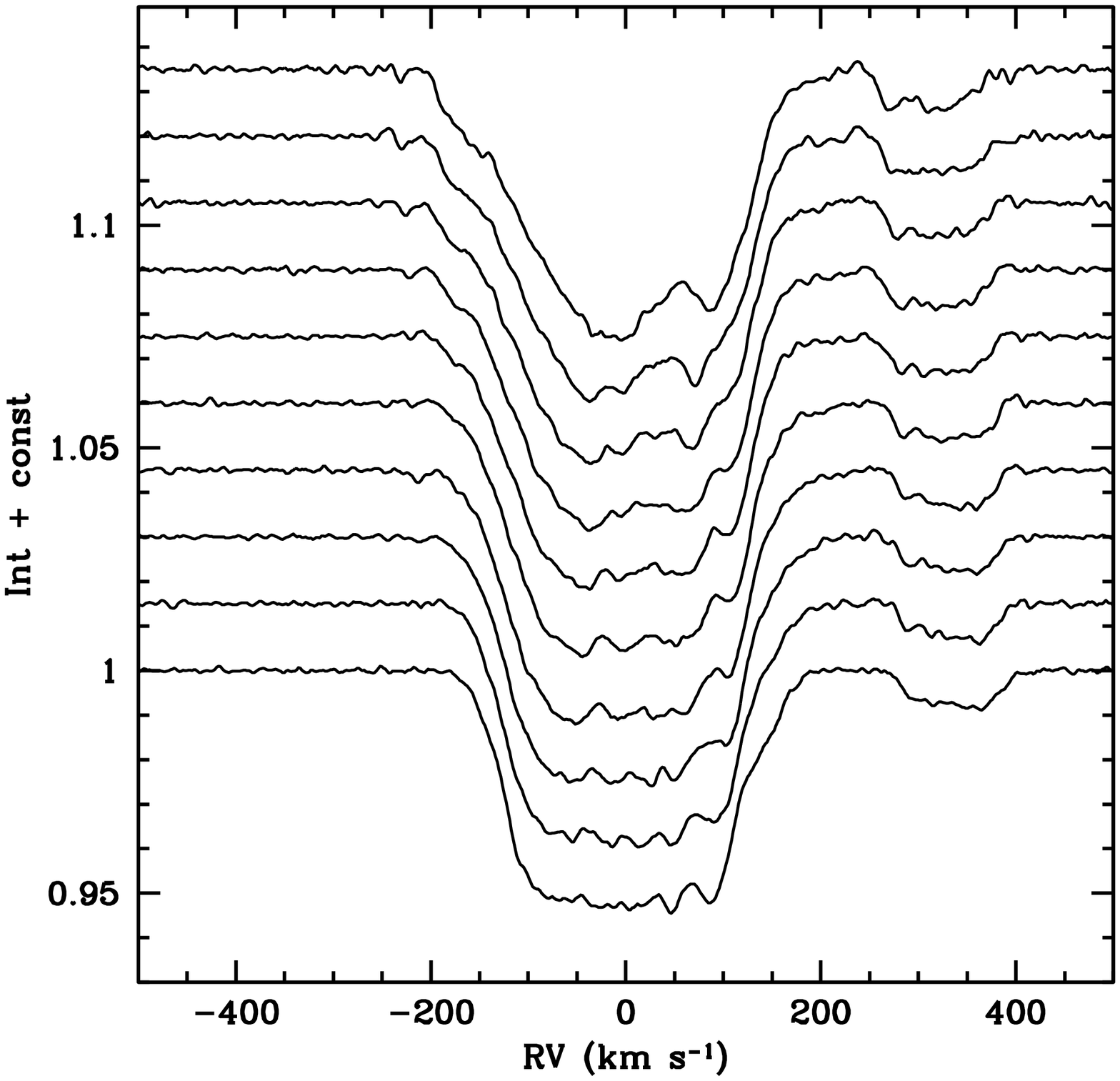}
\includegraphics[width=1.02\linewidth]{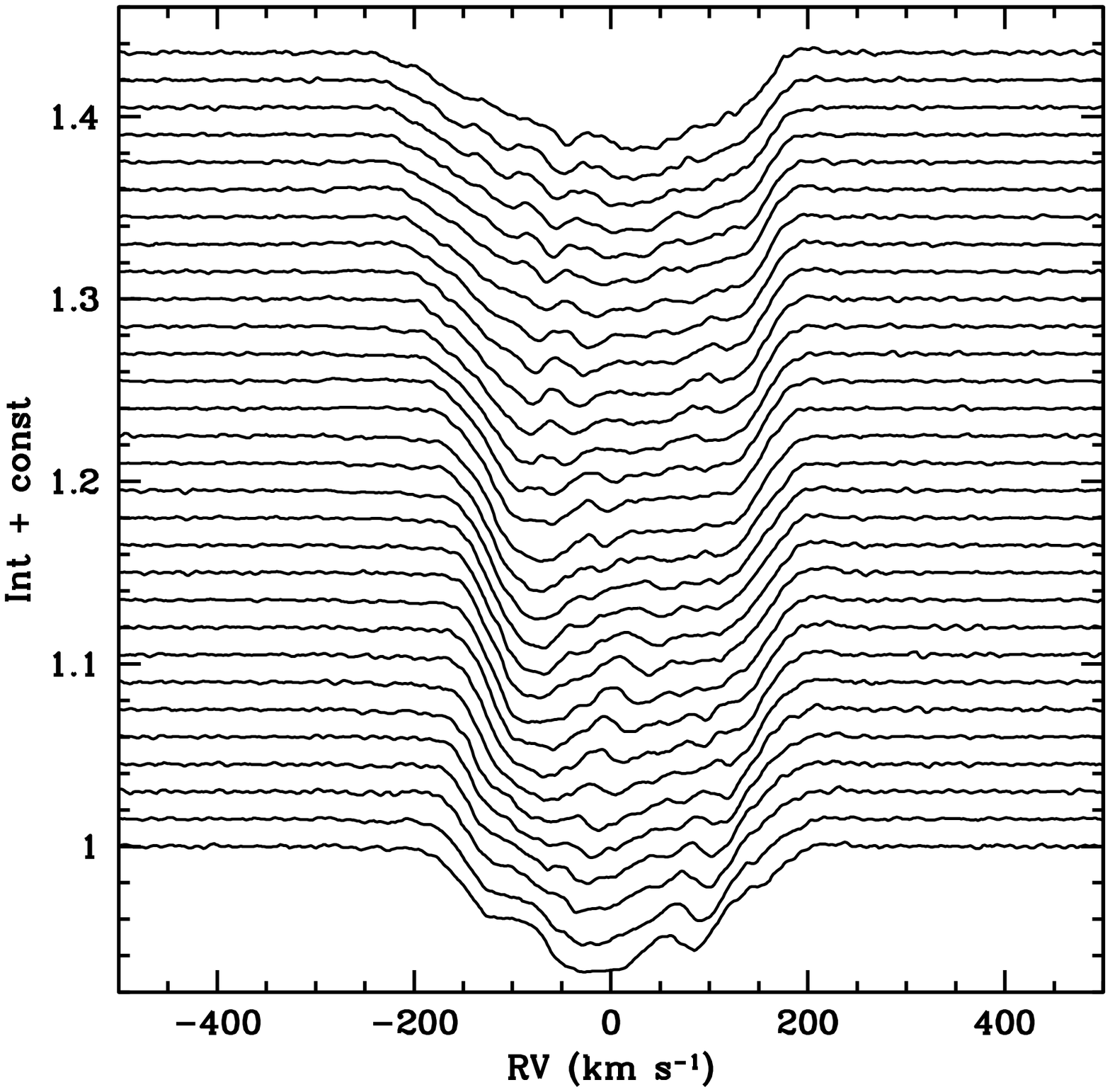}
\caption{\label{averages_block13} {\it Top:} Si III 4552 line profiles obtained on JD 2451653 over a timespan of 3.9 {\rm hrs} (orbital phases 0.890--0.927), showing the emergence of the quasi-centrally-located dip in the profile of $m_1$.  The profile of $m_2$ is located at approximately $+$325 $km~s^{-1}$, and displays the gradual appearance of  a ``blue spike''. Time increases from bottom to top.  {\it Bottom:}  Sequence of profiles obtained on 2451655 over a timespan of 7.5 {\rm hrs} (orbital phases  0.338--0.416).  In this case, the line due to $m_2$ is initially at around $-$25 $km~s^{-1}$ and marches blueward over the course of the night, appearing  near the continuum in the very extended blue wing of $m_1$ by the end of the night.  
}
\end{figure}

\begin{figure} 
\includegraphics[width=1.05\linewidth]{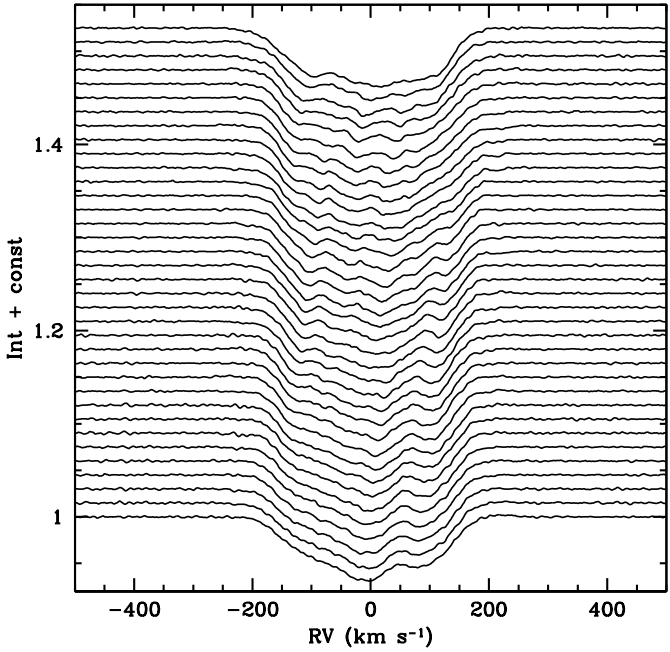}
\caption{\label{averages_block4} Same as previous figure but for the line profiles obtained on JD 2451656 over a timespan of 6.7 {\rm hr} (orbital phases 0.59-0.66).   Noteworthy is the presence of the simultaneous ``blue'' and ``red'' spikes near the time of conjunction ($\phi$0.62).
}
\end{figure}

\begin{figure} 
\includegraphics[width=1.03\linewidth]{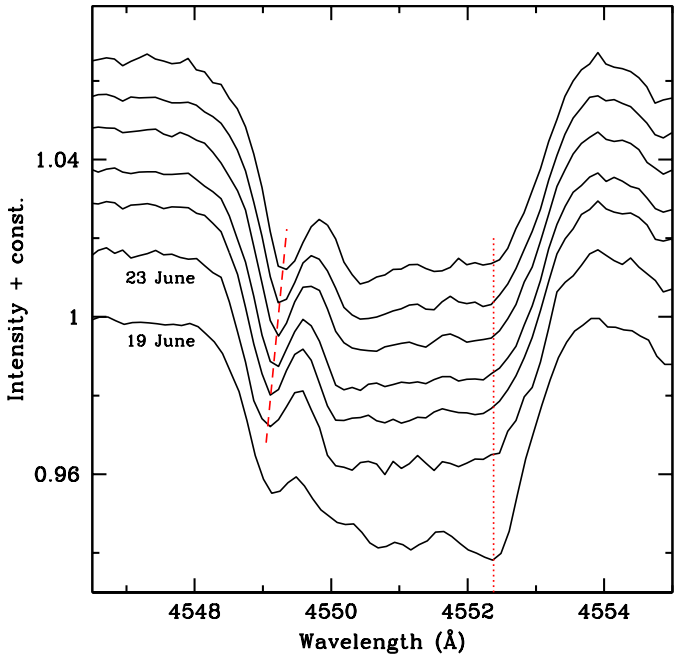}
\caption{Si III 4552 \AA\ absorption line in $m_1$ observed on 2013 June 23 over a 47 min timespan (phases 0.818-0.826), and the  profile of 2013 June 19 (phase 0.822), stacked in order of increasing time from bottom to top.  The prominent blue ``bump'' that migrates from the left absorption wing toward line center is illustrated with the dash line and the quasi-stationary ``red spike'' with the dotted line.  Note the stronger ``red spike'' on the June 19 profile.
\label{2013Jun23_blue_spike}}
\end{figure}

\begin{figure} 
\includegraphics[width=1.03\linewidth]{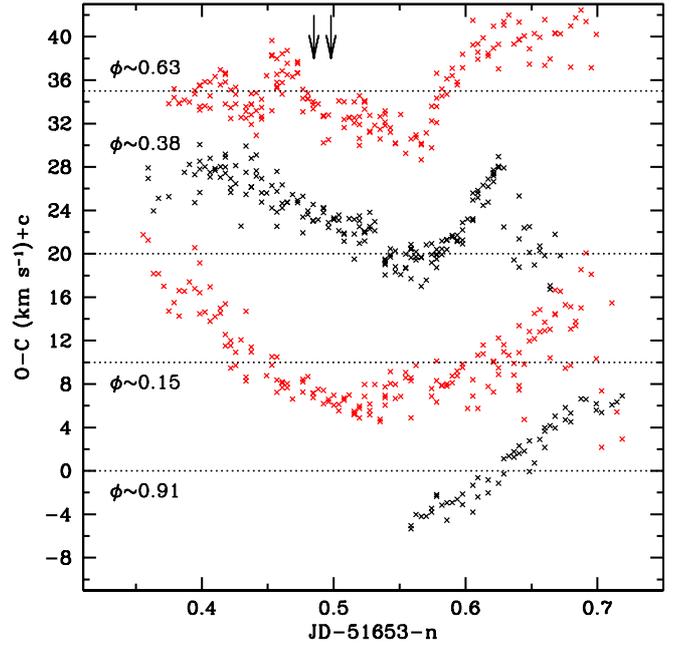}
\caption{Values of O-C from the $m_1$ SOFIN RV curve of Si III 4552 \AA , using the best-fit obtained with FOTEL. From bottom to top, the sets correspond to $JD-2445543-n$, with $n$=0,1,2,3, and each set is shifted along the vertical axis by, respectively, c= 0, 10, 20, 35 km/s, which is indicated by the horizontal lines. The mean orbital phase is indicated, and the arrows point to the two possible times of conjunction when $m_2$ is in back of $m_1$ (see Tables \ref{table_fits_fotel1} and \ref{table_fits_fotel_other}). 
\label{OminusC_m1}}
\end{figure}

\begin{figure} 
\includegraphics[width=1.0\linewidth]{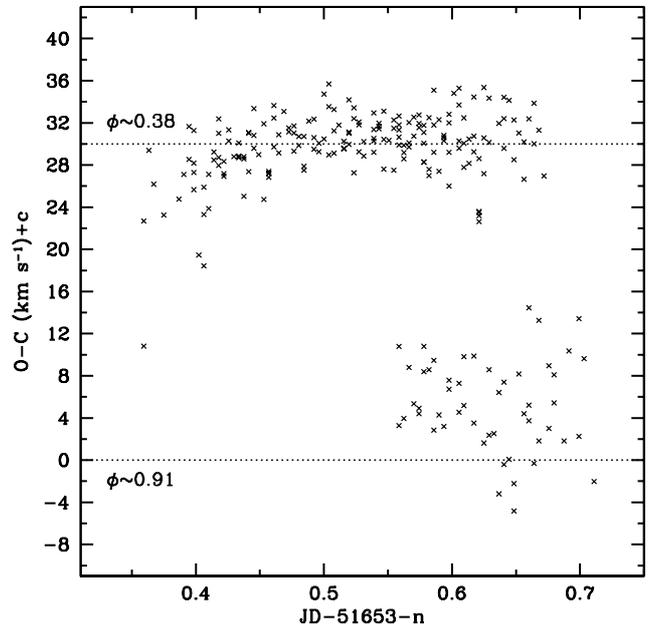}
\caption{Same as above for $m_2$ during the  orbital phases around elongations
\label{OminusC_m2}}
\end{figure}

\begin{table}
\begin{center}
\begin{tiny}
\caption{ESPaDOnS RVs  $m_1$\label{table_RVs_m1}}
\begin{tabular}{ccccccc}
\hline
\hline
{\bf JD$^1$} & {\bf Phase} &{\bf 4552\AA} & {\bf 6678\AA}  & {\bf 4661\AA} & {\bf 4907\AA} & {\bf Comment$^2$} \\
\hline
   54541.081 &    0.146 &   102.0 &    98.8 &    96.8 &    90.4    &   \\   
   54542.002 &    0.376 &    97.3 &    85.2 &    90.3 &    88.8    & bl \\  
   54543.030 &    0.632 &   -53.5 &   -39.4 &   -58.5 &   -62.8    & bl \\   
   54544.853 &    0.086 &    49.1 &    40.4 &    51.8 &    51.6    & bl \\   
   54546.024 &    0.378 &    90.5 &    93.1 &    96.3 &    94.1    &   \\   
   54547.184 &    0.667 &   -58.6 &   -48.4 &   -76.5 &   -70.6    & bl \\   
   54548.031 &    0.877 &  -106.8 &  -108.6 &  -111.3 &  -113.3    &   \\
\hline
\hline
\end{tabular}
\end{tiny}
\end{center}
{\bf Notes:} {\small $^{1}${JD-2400000}; $^2$ bl=blend, indicates that the lines of
$m_1$ and $m_2$ are blended. }
\end{table}

\begin{table}
\begin{center}
\begin{tiny}
\caption{ESPaDOnS RVs $m_2$  \label{table_RVs_m2}}
\begin{tabular}{ccccc}
\hline
\hline
{\bf JD$^1$} & {\bf Phase} &{\bf 4552\AA} & {\bf 6678\AA}  & {\bf 4661 \AA}  \\
\hline
   54541.081 &    0.1463 &  -148.7 &  -151.6 &  -151.1 \\
   54543.030 &    0.6318 &    .... &    .... &    .... \\
   54544.853 &    0.0858 &    .... &    .... &     ....\\
   54546.024 &    0.3776 &  -124.3 &  -127.2 &  -127.5 \\
   54547.184 &    0.6666 &    .... &    .... &    .... \\
   54548.031 &    0.8775 &   168.5 &   178.9 &   165.7 \\
   54549.055 &    0.1326 &  -137.5 &  -139.9 &  -137.1 \\
   54550.098 &    0.3924 &    .... &    .... &    .... \\
   54551.113 &    0.6452 &    .... &    .... &    .... \\
   54551.898 &    0.8408 &   198.3 &   192.0 &   197.2 \\
   54552.984 &    0.1113 &    .... &    .... &    .... \\
   54553.973 &    0.3576 &  -143.4 &  -141.0 &  -140.2 \\
   56430.827 &    0.8652 &   148.5 &   150.3 &   153.2 \\
   56434.852 &    0.8678 &   146.9 &   144.7 &   150.7 \\
   56435.908 &    0.1306 &  -169.3 &  -172.5 &  -173.1 \\
   56436.799 &    0.3527 &    .... &    .... &    .... \\
   56439.935 &    0.1337 &  -179.0 &  -175.8 &  -176.6 \\
   56442.798 &    0.8469 &   166.0 &   171.3 &   162.2 \\
   56462.743 &    0.8150 &   183.6 &   187.7 &    .... \\
   56464.742 &    0.3131 &  -143.3 &  -140.5 &  -139.6 \\
   56466.741 &    0.8110 &   184.4 &   190.7 &    .... \\
   56466.748 &    0.8127 &   184.5 &   188.9 &    .... \\
   56466.755 &    0.8144 &   182.2 &   184.9 &    .... \\
   56466.761 &    0.8160 &   183.0 &   186.1 &   188.2 \\
   56466.768 &    0.8176 &   182.0 &   185.5 &   188.8 \\
   56466.774 &    0.8192 &   182.3 &   185.1 &   180.5 \\
   56470.792 &    0.8199 &   182.2 &   187.0 &    .... \\
   56471.738 &    0.0556 &    .... &    .... &    .... \\
   56472.738 &    0.3047 &  -149.7 &  -140.0 &  -139.5 \\
\hline
\hline
\end{tabular}
\end{tiny}
\end{center}
{\bf Notes:} {\small $^{1}${JD-2400000} }
\end{table}
\clearpage
\vfill\eject

\begin{table}
\begin{center}
\begin{tiny}
\caption{SOFIN measurements \label{table_SOFIN_RVs}}
\begin{tabular}{lllllll}
\hline
\hline
{\bf ID} & {\bf JD}       & {\bf Phase} & {\bf RV1} & {\bf EW1} & {\bf RV2} &{\bf EW2}   \\
       & { $-$2451650}   &           &{\bf $km s^{-1}$}&{\bf $m\AA$} &{\bf $km s^{-1}$}&{\bf $m\AA$}\\
\hline
  23473  &   3.556  &  0.8899  &  -137.  &   199.  &   199.  &    15.  \\
  23475  &   3.561  &  0.8911  &  -137.  &   200.  &   206.  &    13.  \\
  23477  &   3.564  &  0.8918  &  -136.  &   203.  &   199.  &    13.  \\
  23479  &   3.567  &  0.8926  &  -136.  &   211.  &   204.  &    11.  \\
  23480  &   3.570  &  0.8933  &  -136.  &   205.  &   200.  &    12.  \\
  23482  &   3.573  &  0.8941  &  -135.  &   202.  &   199.  &    11.  \\
\hline
\hline
\end{tabular}
\end{tiny}
\end{center}
{\bf Notes:} {\small $^*$Spectra measured for RVs but not used in the line-profile averages due to
poor S/N.}
\end{table}

 \begin{table}
 \begin{center}
 \begin{tiny}
 \caption{SOFIN  measurements}
 \begin{tabular}{lllllll}
 \hline
 \hline
 {\bf ID} & {\bf JD}       & {\bf Phase} & {\bf RV1} & {\bf EW1} & {\bf RV2} &{\bf EW2}   \\
        & { $-$2451650}   &           &{\bf $km s^{-1}$}&{\bf $m\AA$} &{\bf $km s^{-1}$}&{\bf $m\AA$}\\
\hline      
  23647  &   4.408  &  0.1021  &    11.  &   229.  &    -5.  &   ....  \\
  23648  &   4.410  &  0.1026  &     8.  &   226.  &   -14.  &   ....  \\
  23649  &   4.412  &  0.1031  &    10.  &   225.  &   ....  &   ....  \\
  23651  &   4.415  &  0.1038  &    10.  &   230.  &   ....  &   ....  \\
  23653  &   4.417  &  0.1043  &    11.  &   233.  &   ....  &   ....  \\
  23654  &   4.419  &  0.1048  &     9.  &   228.  &   ....  &   ....  \\
\hline
\hline
\end{tabular}
\end{tiny}
\end{center}
\end{table}

 \begin{table}
 \begin{center}
 \begin{tiny}
 \caption{SOFIN  measurements}
 \begin{tabular}{lllllll}
 \hline
 \hline
 {\bf ID} & {\bf JD}       & {\bf Phase} & {\bf RV1} & {\bf EW1} & {\bf RV2} &{\bf EW2}   \\
        & { $-$2451650}   &           &{\bf $km s^{-1}$}&{\bf $m\AA$} &{\bf $km s^{-1}$}&{\bf $m\AA$}\\
\hline
  23739  &   4.551  &  0.1377  &    30.  &   232.  &   -38.  &   ....  \\
  23740  &   4.553  &  0.1382  &    33.  &   238.  &   -30.  &   ....  \\
  23741  &   4.555  &  0.1387  &    33.  &   233.  &   -30.  &   ....  \\
  23742  &   4.557  &  0.1392  &    33.  &   232.  &   -40.  &   ....  \\
  23743  &   4.559  &  0.1397  &    30.  &   235.  &   -27.  &   ....  \\
  23744  &   4.560  &  0.1399  &    34.  &   233.  &   -37.  &   ....  \\
\hline
\hline
\end{tabular}
\end{tiny}
\end{center}
\end{table}

 \begin{table}
 \begin{center}
 \begin{tiny}
 \caption{SOFIN  measurements}
 \begin{tabular}{lllllll}
 \hline
 \hline
 {\bf ID} & {\bf JD}       & {\bf Phase} & {\bf RV1} & {\bf EW1} & {\bf RV2} &{\bf EW2}   \\
        & { $-$2451650}   &           &{\bf $km s^{-1}$}&{\bf $m\AA$} &{\bf $km s^{-1}$}&{\bf $m\AA$}\\
\hline
  23830  &   4.695  &  0.1736  &    68.  &   210.  &   ....  &   ....  \\
  23831  &   4.698  &  0.1743  &    60.  &   231.  &   ....  &   ....  \\
  23833  &   4.702  &  0.1753  &    53.  &   212.  &   ....  &   ....  \\
  23834  &   4.705  &  0.1761  &    59.  &   227.  &   ....  &   ....  \\
  23836  &   4.710  &  0.1773  &    68.  &   206.  &   ....  &   ....  \\
  23838  &   4.714  &  0.1783  &    58.  &   210.  &   ....  &   ....  \\
\hline
\hline
\end{tabular}
\end{tiny}
\end{center}
\end{table}

 \begin{table}
 \begin{center}
 \begin{tiny}
 \caption{SOFIN  measurements}
 \begin{tabular}{lllllll}
 \hline
 \hline
 {\bf ID} & {\bf JD}       & {\bf Phase} & {\bf RV1} & {\bf EW1} & {\bf RV2} &{\bf EW2}   \\
        & { $-$2451650}   &           &{\bf $km s^{-1}$}&{\bf $m\AA$} &{\bf $km s^{-1}$}&{\bf $m\AA$}\\
\hline
  23995  &   5.488  &  0.3711  &   109.  &   202.  &  -175.  &    18.  \\
  23996  &   5.489  &  0.3713  &   109.  &   207.  &  -175.  &    18.  \\
  23997  &   5.491  &  0.3718  &   109.  &   206.  &  -170.  &    17.  \\
  23998  &   5.492  &  0.3721  &   110.  &   211.  &  -172.  &    16.  \\
  24000  &   5.494  &  0.3726  &   110.  &   208.  &  -170.  &    17.  \\
  24001  &   5.495  &  0.3728  &   110.  &   206.  &  -173.  &    17.  \\
\hline
\hline
\end{tabular}
\end{tiny}
\end{center}
\end{table}

 \begin{table}
 \begin{center}
 \begin{tiny}
 \caption{SOFIN  measurements}
 \begin{tabular}{lllllll}
 \hline
 \hline
 {\bf ID} & {\bf JD}       & {\bf Phase} & {\bf RV1} & {\bf EW1} & {\bf RV2} &{\bf EW2}   \\
        & { $-$2451650}   &           &{\bf $km s^{-1}$}&{\bf $m\AA$} &{\bf $km s^{-1}$}&{\bf $m\AA$}\\
\hline
  24079  &   5.586  &  0.3955  &   102.  &   210.  &  -165.  &    20.  \\
  24080  &   5.587  &  0.3958  &   102.  &   203.  &  -163.  &    17.  \\
  24081  &   5.588  &  0.3960  &   102.  &   213.  &  -159.  &    21.  \\
  24084  &   5.590  &  0.3965  &   102.  &   207.  &  -162.  &    15.  \\
  24085  &   5.591  &  0.3968  &   103.  &   211.  &  -167.  &    19.  \\
  24086  &   5.592  &  0.3970  &   102.  &   206.  &  -163.  &    17.  \\
\hline
\hline
\end{tabular}
\end{tiny}
\end{center}
\end{table}

 \begin{table}
 \begin{center}
 \begin{tiny}
 \caption{SOFIN  measurements}
 \begin{tabular}{lllllll}
 \hline
 \hline
 {\bf ID} & {\bf JD}       & {\bf Phase} & {\bf RV1} & {\bf EW1} & {\bf RV2} &{\bf EW2}   \\
        & { $-$2451650}   &           &{\bf $km s^{-1}$}&{\bf $m\AA$} &{\bf $km s^{-1}$}&{\bf $m\AA$}\\
\hline
  24271  &   6.416  &  0.6023  &     3.  &   241.  &     2.  &   ....  \\
  24274  &   6.419  &  0.6030  &     0.  &   232.  &    -5.  &   ....  \\
  24275  &   6.420  &  0.6032  &     3.  &   246.  &    -5.  &   ....  \\
  24276  &   6.421  &  0.6035  &     3.  &   243.  &     1.  &   ....  \\
  24277  &   6.422  &  0.6037  &     1.  &   240.  &    -1.  &   ....  \\
  24278  &   6.423  &  0.6040  &    -1.  &   234.  &   ....  &   ....  \\
\hline
\hline
\end{tabular}
\end{tiny}
\end{center}
\end{table}

 \begin{table}
 \begin{center}
 \begin{tiny}
 \caption{SOFIN  measurements}
 \begin{tabular}{lllllll}
 \hline
 \hline
 {\bf ID} & {\bf JD}       & {\bf Phase} & {\bf RV1} & {\bf EW1} & {\bf RV2} &{\bf EW2}   \\
        & { $-$2451650}   &           &{\bf $km s^{-1}$}&{\bf $m\AA$} &{\bf $km s^{-1}$}&{\bf $m\AA$}\\
\hline
  24359  &   6.514  &  0.6267  &   -16.  &   238.  &   ....  &   ....  \\
  24360  &   6.516  &  0.6272  &   -15.  &   238.  &   ....  &   ....  \\
  24361  &   6.517  &  0.6274  &   -19.  &   240.  &   ....  &   ....  \\
  24364  &   6.519  &  0.6279  &   -17.  &   253.  &   ....  &   ....  \\
  24365  &   6.520  &  0.6282  &   -18.  &   237.  &   ....  &   ....  \\
  24366  &   6.521  &  0.6284  &   -20.  &   242.  &   ....  &   ....  \\
\hline
\hline
\end{tabular}
\end{tiny}
\end{center}
\end{table}

 \begin{table}
 \begin{center}
 \begin{tiny}
 \caption{SOFIN  measurements}
 \begin{tabular}{lllllll}
 \hline
 \hline
 {\bf ID} & {\bf JD}       & {\bf Phase} & {\bf RV1} & {\bf EW1} & {\bf RV2} &{\bf EW2}   \\
        & { $-$2451650}   &           &{\bf $km s^{-1}$}&{\bf $m\AA$} &{\bf $km s^{-1}$}&{\bf $m\AA$}\\
\hline
  24444  &   6.648  &  0.6600  &   -32.  &   231.  &   ....  &   ....  \\
  24445  &   6.651  &  0.6608  &   -31.  &   239.  &   ....  &   ....  \\
  24448  &   6.655  &  0.6618  &   -32.  &   233.  &   ....  &   ....  \\
  24449  &   6.657  &  0.6623  &   -31.  &   244.  &   ....  &   ....  \\
  24450  &   6.660  &  0.6630  &   -36.  &   242.  &   ....  &   ....  \\
  24451  &   6.663  &  0.6638  &   -36.  &   244.  &   ....  &   ....  \\
  24453  &   6.666  &  0.6645  &   -34.  &   250.  &   ....  &   ....  \\
  24454  &   6.668  &  0.6650  &   -34.  &   227.  &   ....  &   ....  \\
  24455  &   6.671  &  0.6658  &   -34.  &   236.  &   ....  &   ....  \\
  24458  &   6.676  &  0.6670  &   -38.  &   247.  &   ....  &   ....  \\
  24459  &   6.679  &  0.6678  &   -35.  &   234.  &   ....  &   ....  \\
  24461  &   6.683  &  0.6688  &   -35.  &   237.  &   ....  &   ....  \\
  24462  &   6.686  &  0.6695  &   -35.  &   240.  &   ....  &   ....  \\
  24464  &   6.690  &  0.6705  &   -36.  &   242.  &   ....  &   ....  \\
  24466  &   6.694  &  0.6715  &   -41.  &   248.  &   ....  &   ....  \\
  24468  &   6.698  &  0.6725  &   -39.  &   256.  &   ....  &   ....  \\
  24470  &   6.703  &  0.6737  &   -37.  &   257.  &   ....  &   ....  \\
\hline
\hline
\end{tabular}
\end{tiny}
\end{center}
\end{table}

\end{appendix}
\end{document}